\documentclass[%
 aip,
 amsmath,amssymb,
 reprint,%
]{revtex4-1}

\usepackage{graphicx}
\usepackage{dcolumn}
\usepackage{bm}

\usepackage[utf8]{inputenc}
\usepackage[T1]{fontenc}
\usepackage{mathptmx}

\usepackage{hyperref,color,subcaption}
\usepackage[normalem]{ulem}		

\hypersetup{
	colorlinks=true,    
	linkcolor=blue,
	citecolor=blue,
	urlcolor=blue,
	bookmarks=true,
	pdfusetitle=true   
}

\definecolor{corr}{rgb}{0,0,0}

\newif\ifshowsout
\showsoutfalse
\newcommand{\showsout}[1]{\ifshowsout#1\fi}

\begin{document}

\preprint{AIP/123-QED}

\title[]{Kinetic theory based solutions for particle clustering in turbulent flows}

\author{C.P.~Stafford}
\affiliation{Department of Chemical and Biological Engineering, Iowa State University, Ames, IA 50011, USA}
\email{cs9@iastate.edu}
\author{D.C.~Swailes}%
\affiliation{School of Mathematics, Statistics \& Physics, Newcastle University, Newcastle upon Tyne NE1 7RU, UK}%
\author{M.W.~Reeks}%
\affiliation{School of Engineering, Newcastle University, Newcastle upon Tyne NE1 7RU, UK}%

\date{\today}

\begin{abstract}
Kinetic theory provides an elegant framework for studying dispersed particles in turbulent flows. Here the application of such probability density function (PDF)-based descriptions is considered in the context of particle clustering. The approach provides a continuum representation for the particle phase in which momentum conservation identifies two fundamental contributions to the particle mass flux. These take the form of an additional body force, which emerges from inhomogeneities in the sampling of turbulence by particles, and a component of the particle phase stress tensor associated with turbophoresis. Remarkably, these contributions are frequently overlooked in the specification of mean-field models for dispersed particle flows. To assess the relative importance of these mass flux contributions a kinematic simulation study has been performed, making use of a specially constructed inhomogeneous flow field designed to mimic the dynamics of particle pair behaviour. Whilst the turbophoretic contribution always acts to increase the clustering of particles, both the direction in which the additional body force acts and the relative importance of the two mass flux contributions are found to vary with particle inertia and turbulence intensity. In some regimes the body force is also the dominant mechanism responsible for particle clustering, demonstrating its importance within model formulations. It is further highlighted that the evolution of the radial distribution function which describes particle clustering is represented as a balance between convection and diffusion, and only inclusion of the identified mass flux contributions within this balance enables correct prediction of the particle pair concentration profile observed in simulations.
\end{abstract}

\maketitle

\section{Introduction} \label{sec:intro}

The clustering of particles in turbulent flow is a striking feature of dispersed particle transport, and has been subject to extensive investigation \cite{Balachandar2010}. Despite this, a definitive classification of this important phenomenon is still not available, with many questions yet to be answered. Identification of the key physical mechanisms responsible for this effect is a focus of ongoing research, and requires consideration of how structures within the dispersed phase are affected by those within the carrier phase. Such research is essential for improving the understanding of situations in which turbulence effects are important, including particle dispersion, deposition, and agglomeration in boundary layer flows. These physical processes occur in a multiplicity of natural and engineered environments, such as cloud formation in the atmosphere, particle transport in pipelines, spray combustion of fuel droplets, and the atmospheric dispersion of emissions. Of particular utility is the ability to predict the clustering behaviour of particle-laden flows through the use of modelling approaches \cite{Reeks2021}, with the description provided by such methods offering insight into the multiscale behaviour that affects how these systems evolve.

Within the realm of modelling dispersed particle flows, probability density function (PDF) methods offer a comprehensive means of capturing the effects of turbulence on the particle behaviour. In particular, the kinetic approach pioneered by Reeks \cite{Reeks2021} is suitable for describing the transport of small inertial particles in a manner analogous to the kinetic theory of gases. Treatment of the turbulent flow as a spatio-temporally correlated stochastic field in this procedure has the consequence that the terms which emerge in the resultant kinetic equation are then able to describe in detail the collective particle motion induced by the turbulence. A previous study \cite{Stafford2021} has focused upon the ability of the kinetic PDF approach at capturing the enhancement in gravitational settling experienced by small inertial particles; the present work utilises this kinetic framework as a means of describing particle clustering in turbulence. {\color{corr} This is advantageous since the terms present in the kinetic PDF approach are able to account for the various mechanisms which contribute to particle clustering. Importantly, this framework contains additional terms that are crucial to capturing the correct clustering behaviour, in contrast to other modelling approaches in which these terms do not appear.}

\subsection{Clustering mechanisms} \label{sec:mechanisms}

{\color{corr} Whilst turbulence-induced clustering of particles has become a recognised phenomenon, many different, potentially competing, explanations have been proposed to explain and quantify this behaviour.} An early observation by Reeks \cite{Reeks1983} noted the tendency for particles to migrate in the direction of decreasing turbulence level, a phenomenon that has been termed \emph{turbophoresis}, and that arises from a force balance between the net drag force and the gradient of the turbulence induced particle kinetic stresses. {\color{corr}Such an effect naturally contributes to the non-uniformity of the spatial particle distribution, and is intrinsic within many models of particle-laden flow. There are various other aspects of the carrier flow and particle behaviour which need to be accounted for within a modelling framework in order to describe particle clustering.}

{\color{corr}A key determinant of particle clustering is the spatial structure of the carrier flow.} One of the first works that revealed the clustering of particles in turbulence was carried out by Maxey in the context of gravitational settling \cite{Maxey1987,Maxey1987a}, with the main finding of an enhancement in the average settling velocity arising as a consequence of the preferential concentration of particles due to the influence of the structures in the flow field. It was demonstrated that for low inertia particles, the divergence of the effective particle velocity field is positive in regions of high vorticity, and negative in regions of high strain rate. This results in an inertial bias on the particles, causing them to accumulate in regions of high strain rate, and thus leading to locally non-uniform particle concentrations{\color{corr}; an effect referred to as \emph{centrifuging}}.

Direct numerical simulation (DNS) studies focusing upon particle clustering were subsequently undertaken by Squires \& Eaton \cite{Squires1991a}, identifying conditions under which preferential concentration was most pronounced. This showed that the turbulence could, in some cases, inhibit rather than enhance the mixing of particles, going against the conventional view that considered particle mixing as a homogenizing process resulting in uniform distributions. {\color{corr}Additionally, this work also identified centrifuging} as being the mechanism responsible for clustering of low-inertia particles even in the absence of gravity.
\showsout{\sout{This early numerical work on particle clustering set the scene for more detailed investigations into the mechanisms responsible for this behaviour.}}

\showsout{\sout{Another phenomenon associated with clustering is}} {\color{corr}Another effect arising from the carrier flow structure is} the \textit{sweep-stick mechanism} \cite{CHEN2006,Coleman2009}, which stems from the observation that clustering of particles in homogeneous isotropic turbulence (HIT) correlates with stagnation points in the acceleration field of the carrier fluid. This \showsout{\sout{correlation results from}} {\color{corr}is due to} these regions of zero-acceleration in the fluid being swept together with inertial particles by the large scale motions of the fluid. When the position of a particle is co-incident with an acceleration stagnation point, its subsequent motion aligns with the fluid velocity for longer than would otherwise be expected. This mechanism is applicable to particles with a small inertia, and is dependent upon the dynamic sweeping of small scale motions by the larger scales within the turbulence \cite{Monchaux2012}. The result is spatial clustering of particles over multiple different scales, with the sweep-stick mechanism reported to be dominant over the preferential concentration arising from centrifuging out of high vorticity regions \cite{Coleman2009}.

\showsout{\sout{More recent work has focused on the mechanisms responsible for clustering}} {\color{corr}The multi-valued nature of inertial particle trajectories in velocity space also has implications for particle clustering.} Outside of the low-inertia {\color{corr}regime, this is seen through} the development of \textit{caustics} \cite{Wilkinson2005}, in which multiple point-particles are deemed to occupy the same spatial location but with differing velocities.
The formation of caustics implies spatial regions with greatly increased particle concentrations, with consequently greater rates of collisions and coagulation. Clustering is therefore associated with a network of caustic lines. \showsout{\sout{This mechanism holds particular relevance for explaining the rapid onset of rainfall from convecting clouds \cite{Wilkinson2006}. The influence of particle relative velocities on spatial distribution and collision rates has been investigated in a number of works \cite{Gustavsson2014a,Bragg2014,Perrin2015,Bhatnagar2018}, with the importance of caustics in this process being highlighted. It is known that the collision rate is affected by both the radial distribution function \cite{Reade2000} and the mean radial relative velocities \cite{WANG2000}, and it is reported that the dependence of the latter on Stokes number is attributable to the formation of caustics \cite{Perrin2015}.}}

Several studies on clustering have highlighted 
the role of the velocity gradient tensor sampled by particles \cite{Zaichik2007,Bragg2014a}.
Bragg \textit{et al}. \cite{Bragg2015} {\color{corr} carried out} a detailed analysis of the mechanisms responsible for particle clustering in the inertial range of isotropic turbulence to determine their relative contribution to the net drift.
\showsout{\sout{In terms of the Kolmogorov scaled Stokes number $St_{\eta}$,}} It was found that for \showsout{\sout{$St_{\eta} \ll 1$}} 
{\color{corr}very low-inertia particles,} clustering is due to preferential sampling of the coarse-grained fluid velocity gradient tensor, consistent {\color{corr}with the centrifuging explanation} of Maxey \cite{Maxey1987a}.
In contrast, for \showsout{\sout{$St_{\eta} > 1$}} {\color{corr}high-inertia particles} a clustering mechanism referred to as the \textit{non-local path history effect} was identified{\color{corr}, which} involves the sampling of larger fluid velocity differences along particle trajectories.
Further investigation\cite{Bragg2015a} found that, despite the dominant clustering mechanism changing from centrifuging to the symmetry breaking effects of the non-local path history {\color{corr}as the particle inertia is increased}, the particle positions \showsout{\sout{continue to}}  correlate with high strain, low rotation regions of turbulence in both cases. This is due to the path history effects being influenced by the preferential sampling of the fluid velocity gradient tensor along the particle trajectories in such a way as to generate a bias for clustering in high strain regions of the turbulence. \showsout{\sout{Over the timescale for which particles retain memory of their path history, it was observed that the strain and rotation fields \showsout{\sout{along the particle trajectory}} {\color{corr}sampled by particles} remain significantly correlated, with this preferential sampling affecting the non-local contributions to the drift velocity, thereby subsequently generating the clustering.}} {\color{corr} It is worth noting the common Lagrangian aspect of both the non-local path history and the sweeping part of the sweep-stick mechanism, and possible shared preferential sampling of velocity gradients as a result. However, it remains to be clarified how interdependent these two mechanisms are.}

Given the variety of \showsout{\sout{physical}} mechanisms which have been proposed to explain and quantify \showsout{\sout{the phenomenon of}} particle clustering, it is clear that the ability to capture and represent such effects within continuum models for dispersed particle flows is important. Such a representation therefore needs to respect the multi-valued nature of inertial particle trajectories in \showsout{\sout{configuration}} {\color{corr}velocity} space\showsout{\sout{ to account for caustic effects}}, as well as incorporate \showsout{\sout{physical}} information about both the spatial structure of the carrier flow and path history of the velocity gradient tensor \showsout{\sout{along particle trajectories to capture the path history effects}}. To \showsout{\sout{that end}} {\color{corr} address this requirement, the present work} focuses on the use of the kinetic PDF approach as a means of constructing suitable descriptions.

\subsection{Particle pair models} \label{sec:partpair}

In this work the statistical analysis of particle clustering is carried out through the use of particle pair models, in which the relative separation between particles is studied. Previous work on such models has identified the radial distribution function (RDF) as a key measure of particle clustering \cite{Rani2014}. The first analytical model for the RDF was obtained by Zaichik \& Alipchenkov \cite{Zaichik2003}, who interpreted the clustering phenomenon in homogeneous turbulence as a result of a particle migration drift due to the gradient of the radial relative fluctuating velocity, and therefore identifying this drift as an additional attractive velocity. This work involved the development of a kinetic equation for the joint PDF for particle pair separations and relative velocities,
with the Furutsu-Novikov formula being used to analyse the unclosed terms.
At small Stokes number $St$, this lead to the RDF displaying power law behaviour with an exponent that is proportional to $St^2$, providing a benchmark for quantifying particle clustering.

Further work by Zaichik \& Alipchenkov \cite{Zaichik2007} refined this approach using improved closure methodologies.
Their earlier work \cite{Zaichik2003} modelled the two-point Lagrangian fluid velocity structure functions 
by assuming equal timescales for the Lagrangian two-time strain and rotation correlations.
However, Brunk \textit{et al}.~ \cite{Brunk1997} observed through use of the DNS data of Girimaji \& Pope \cite{Girimaji1990} that these timescales can differ significantly.
By modelling the structure functions in terms of the strain and rotation rate tensors, the different timescales introduced by these components ensures the turbulence-particle interaction mechanism that contributes to the preferential clustering is captured to a greater extent. Extension of the applicability of this work from zero-size to finite-size particles was also considered by Zaichik \& Alipchenkov through applying boundary conditions to the continuum equations at a separation equal to the particle diameter \cite{Zaichik2009}.

In another pioneering study Chun \textit{et al}. \cite{CHUN2005} developed an analytical theory for predicting the
RDF for small particles in turbulent flows. This supposed that the preferential concentration of particles at lengthscales smaller than the Kolmogorov scale can be attributed to a radial inward drift of inertial particles in a locally linear flow field. This analysis was performed within a particle pair framework by means of a perturbation expansion on the particle equations of motion written in terms of the particle separation and relative velocity. A PDF equation for the distribution of relative particle separation was derived and  solved using the method of characteristics, yielding expressions for the drift and diffusive fluxes experienced by particle pairs. The resultant form of the RDF exhibits the expected power law behaviour with an exponent that is proportional to $St^2$, in agreement with Zaichik \& Alipchenkov \cite{Zaichik2003}.

Further analysis of the approaches of Chun \textit{et al}.~and Zaichik \& Alipchenkov was undertaken by
Bragg \& Collins \cite{Bragg2014a} in order to compare the two theoretical models and gauge their applicability.
It was observed that the model proposed by Chun \textit{et al}.~neglects the effect of the path history contribution due to truncation of the perturbation expansion used, with the physical explanation of the drift coming purely from the centrifuge mechanism of oversampling strain over rotation. Additionally, the effect of caustics is not taken into account. Nonetheless, whilst subject to the restriction of $St \ll 1$, the Chun \textit{et al}.~model does still capture the leading order effects of the clustering mechanism behaviour in the Zaichik \& Alipchenkov model.
However, it is incorrectly assumed that the particle relative velocity field is incompressible, meaning that the Chun \textit{et al}.~model is unable to describe the non-centrifuge path history mechanism proposed by Ijzermans \textit{et al}.~\cite{IJZERMANS2010} involving evolution of the Jacobian tensor along trajectories. In contrast, the Zaichik \& Alipchenkov model captures the path history contribution, whilst also including the effect of caustics, although despite incorporating these mechanisms a quantitative discrepancy in the power law exponent of the RDF is exhibited.

The respective shortcomings of these existing particle pair models highlights the need for more comprehensive approaches and closures which are able to better incorporate the causal mechanisms of clustering. The present work addresses this need by using a kinetic equation to describe the two-particle separation in the same vein as Zaichik \& Alipchenkov, but with the utilisation of a more detailed closure strategy to better capture the physics of the clustering behaviour.

The remainder of the paper is structured as follows. Section~\ref{sec:kinetic-pdf-approach} introduces the kinetic approach for a two-particle description, and reviews the development of particle pair models for describing particle clustering and the role of the current work within this. Section~\ref{sec:inhomogeneous-flow-model} details an inhomogeneous flow field which is used to investigate the behaviour of particle pairs in this work.
Section \ref{sec:LHA} considers an existing simple closure for the terms involved in the kinetic approach, and the implications arising from this. Section \ref{sec:numerics} outlines the numerical procedures used to simulate the flow field and evaluate the relevant particle-phase statistics for the kinetic approach. Section \ref{sec:results} analyses the results obtained from the numerical simulations in terms of the particle concentration and relevant contributions towards the enhancement or suppression of clustering in a particle pair framework. Section \ref{sec:conclusions} summarises the findings of this work, and discusses the implications on developing closure models for particle pair behaviour.

\section{Kinetic PDF approach description} \label{sec:kinetic-pdf-approach}

\subsection{Two-particle formulation} \label{sec:PDF-2p}

The standard one-particle PDF approach is based on consideration of a particle equation of motion, which in general form for particles with positions $\bm{x}_p(t)$ and velocities $\bm{v}_p(t)$ can be expressed as
\begin{equation} \label{eq:part-eq}
    \ddot{\bm{x}}_p = \bm{F}(\bm{x}_p,\bm{v}_p,t) + \bm{f}(\bm{x}_p,t) \, ,
    \qquad
    \begin{array}{ll}
        \bm{x}_p(t_0) & \hspace{-2mm} = \bm{x}^{0} \\
        \dot{\bm{x}}_p(t_0) & \hspace{-2mm} = \bm{v}^{0}
    \end{array}
    \, ,
\end{equation}
where $\bm{F}(\bm{x},\bm{v},t)$ defines the mean force per unit mass acting on particles with position $\bm{x}$ and velocity $\bm{v}$ at time $t$, and $\bm{f}(\bm{x},t)$ is a zero-mean stochastic field modelling the fluctuating acceleration experienced by a particle with position $\bm{x}$ at time $t$. The notation convention used here is that $(\bm{x},\bm{v})$ denotes coordinates in phase-space, whilst $(\bm{x}_p,\bm{v}_p)$ denotes the particle phase-space position at time $t$. This work focuses upon the specific case of a linear drag law for the particle motion, which corresponds to the case
\begin{subequations}
    \label{eq:linear-drag}
    \begin{align}
        \bm{F}(\bm{x},\bm{v},t) & = \beta (\langle \bm{U}(\bm{x},t) \rangle - \bm{v}) \, , \label{eq:mean-force} \\
        \bm{f}(\bm{x},t) & = \beta \bm{U}^{\prime}(\bm{x},t) \, ,
    \end{align}
\end{subequations}
where $\beta = \tau_p^{-1}$ denotes the particle inertia parameter, and $\bm{U} = \langle \bm{U} \rangle + \bm{U}^{\prime}$ is the Reynolds decomposition of the turbulent carrier flow velocity into mean and fluctuating parts respectively.

To develop a two-particle description, monodisperse particles with trajectories $\bm{x}_p^1$, $\bm{x}_p^2$, and velocities $\bm{v}_p^1$, $\bm{v}_p^2$ are considered. The particle separation $\bm{r}_p = \bm{x}_p^2 - \bm{x}_p^1$ and relative velocity $\bm{w}_p = \bm{v}_p^2 - \bm{v}_p^1$ then define a particle pair  description in a Lagrangian frame of reference centred on the target particle $\bm{x}_p^1$. For particles governed by the linear drag law in Eqs.~\eqref{eq:part-eq} and  \eqref{eq:linear-drag}, the evolution of the two-particle system is then described by
\begin{equation} \label{eq:two-particle-eq}
    \ddot{\bm{r}}_p = \beta \left( \bm{U} (\bm{r}_p + \bm{x}_p^1,t) - \bm{U} (\bm{x}_p^1,t) - \bm{w}_p \right) \, ,
    \qquad
    \begin{array}{ll}
        \bm{r}_p(t_0) & \hspace{-2mm} = \bm{r}^0 \\
    \bm{w}_p(t_0) & \hspace{-2mm} = \bm{w}^0
    \end{array}
    \, .
\end{equation}
The relative velocity field
\begin{displaymath}
\Delta \bm{U}(\bm{r},t;\bm{x}) =
\bm{U} (\bm{r} + \bm{x},t) - \bm{U} (\bm{x},t) 
\end{displaymath}
will clearly be inhomogeneous even when, as assumed, the underlying field $\bm{U}$ is homogeneous. However, when $\bm{U}$ is homogeneous the statistics of $\Delta \bm{U}$ will be independent of $\bm{x}$. Following the approach implicit in the work of Zaichik \& Alipchenkov \cite{Zaichik2003,Zaichik2007,Zaichik2009,zaichik2008particles}, this motivates
the construction of an inhomogeneous fluid velocity field
\begin{equation} \label{eq:inhomogeneous-flu-vel}
\bm{u}(\bm{r},t) = \Delta \bm{U} (\bm{r},t;0) = \bm{U}(\bm{r},t) - \bm{U}(\bm{0},t) \, .
\end{equation}
as a proxy for the relative velocity field. Note that $\bm{u}(\bm{0},t) = \bm{0}$, so in a one-particle context the flow field can be interpreted as an effective boundary layer. 

The resulting model for describing the joint distribution of $(\bm{r}_p,\bm{w}_p)$ will be exact provided the sampling of $\bm{U}$ by $\bm{x}_p^1$ is unbiased. Notwithstanding this Eulerian approximation, the model given by Eq.~\eqref{eq:inhomogeneous-flu-vel} affords the simple construction of an appropriate inhomogeneous flow field in which relative displacements $\bm{r}_p$ and velocity $\bm{w}_p$ are governed by
\begin{equation} \label{eq:two-particle-eq-approx}
\ddot{\bm{r}}_p = \beta \left( \bm{u}(\bm{r}_p,t) - \bm{w}_p \right) \, .
\end{equation}
Eq.~\eqref{eq:two-particle-eq-approx} takes the same form as the single particle model given by Eq.~\eqref{eq:part-eq}, meaning that the associated two-particle PDF $p(\bm{r},\bm{w},t)$ which describes the joint distribution of $\bm{r}_p$ and $\bm{w}_p$ can be interpreted directly within the existing framework for the usual one-particle PDF $p(\bm{x},\bm{v},t)$ for monodisperse particles. The two-particle PDF is therefore governed by a kinetic equation which has the same underlying form as that for single particles, with these different interpretations dealing with the large and small scales of the turbulence respectively\cite{Reeks2021}. The kinetic equation can take various forms depending on how the turbulent flux term is closed \cite{Bragg2012}; for the specific case in which the functional correlation splitting procedure of Furutsu \cite{Furutsu1963} and Novikov \cite{Novikov1965} is used, this transport equation can be written as \cite{Reeks1991,Hyland1999a,Swailes1997}
\begin{equation} \label{eq:kinetic-equation}
    \frac{\partial p}{\partial t} = -
\nabla_{\bm{r}}  \cdot \left[ p \bm{w} \right] -
\nabla_{\bm{w}} \cdot
\Bigl [
p\left( \bm{F} + \boldsymbol{\kappa} \right) -
\nabla_{\bm{r}} \cdot \left[ p \boldsymbol{\lambda} \right] -
\nabla_{\bm{w}} \cdot \left[ p \boldsymbol{\mu} \right]
\Bigr ] ,
\end{equation}
where, for this two-particle interpretation,  $\bm{F}$ is defined by
$\bm{F}(\bm{r},\bm{w},t) = \beta (\langle \bm{u}(\bm{r},t) \rangle - \bm{w})$. It should be stressed that Eq.~\eqref{eq:kinetic-equation} makes no assumption on the range of particle inertia for which the closure is valid, and defines $p$ exactly provided only that the stochastic velocity field $\bm{u}$ is Gaussian. Moreover for non-Gaussian fields Eq.~\eqref{eq:kinetic-equation} represents a first-order model approximation in which the higher-order cumulant contributions are neglected. The kinetic dispersion tensor coefficients $\boldsymbol{\kappa}, \boldsymbol{\lambda}, \boldsymbol{\mu}$ appearing in Eq.~\eqref{eq:kinetic-equation} depend on the underlying carrier flow turbulence embodied in the statistical representation of the inhomogeneous flow field $\bm{u}$, and describe its effect on the particle phase. This can be seen more clearly by considering the related set of continuum equations for the particle phase mean-field variables. Specifically, the first three velocity weighted moments of $p$ are the particle number density $\rho$, the mean relative velocity $\overline{\bm{w}}$ and the kinetic stresses $\overline{\bm{c}\bm{c}}$ (where $\bm{c} = \bm{w} - \overline{\bm{w}}$), defined as
\begin{subequations}
    \label{eq:mean-field-var}
    \begin{align}
        \rho & = \int_{\bm{w}} p \, d\bm{w} \, ,\\
        \overline{\bm{w}} & = \frac{1}{\rho} \int_{\bm{w}} \bm{w} p \, d\bm{w} \, ,\\
        \overline{\bm{c}\bm{c}} & = \frac{1}{\rho} \int_{\bm{w}} (\bm{w} - \overline{\bm{w}}) (\bm{w} - \overline{\bm{w}}) p \, d\bm{w} .
    \end{align}
\end{subequations}
The transport equations for these mean-field variables are obtained by taking corresponding relative velocity-weighted integrals of the kinetic equation Eq.~\eqref{eq:kinetic-equation}, and are \cite{Reeks1991,Hyland1999a,Swailes1997}
\begin{subequations}
        \label{eq:continuum-equations}
    \begin{align}
        \frac{D \rho}{D t} & = - \rho \nabla \cdot \overline{\bm{w}} \label{eq:number-density-equation} \, ,\\
        \rho \frac{D \overline{\bm{w}}}{D t} & = - \nabla \cdot \left[ \rho (\overline{\bm{c}\bm{c}} + \overline{\boldsymbol{\lambda}}) \right] + \rho (\overline{\bm{F}} + \overline{\boldsymbol{\kappa}}) \label{eq:momentum-equation} \, ,\\
        \rho \frac{D}{D t} \overline{\bm{c}\bm{c}} & = - \nabla \cdot \left[ \rho \overline{\bm{c}\bm{c}\bm{c}} \right] + \rho (\boldsymbol{\Psi} + \boldsymbol{\Psi}^{\top}) \, , \label{eq:kinetic-stress-equation}
    \end{align}
\end{subequations}
where
\begin{equation} \label{eq:kinetic-stress-psi}
    \boldsymbol{\Psi} = -\frac{1}{\rho} \nabla \cdot \left[ \rho \overline{\boldsymbol{\lambda} \bm{c}} \right] - \left( \overline{\bm{c}\bm{c}} + \overline{\boldsymbol{\lambda}}^{\top} \right) \cdot  \nabla \overline{\bm{w}} + \overline{\boldsymbol{\mu}} + \overline{\bm{c} \left( \bm{F} + \boldsymbol{\kappa} \right)} \, .
\end{equation}
Here
\begin{equation} \label{eq:material-derivative}
    \frac{D}{D t} = \frac{\partial}{\partial t} + \overline{\bm{w}} \cdot \nabla \, ,
\end{equation}
and the density weighted averaged quantities, denoted using an overbar, are defined as
\begin{equation}
    \overline{\bm{F}} = \frac{1}{\rho} \int_{\bm{w}} \bm{F} p \, d\bm{w} \qquad \textit{etc.}
\end{equation}
The appearance of the density weighted dispersion tensors $\overline{\boldsymbol{\kappa}}, \overline{\boldsymbol{\lambda}}, \overline{\boldsymbol{\mu}}$ in Eqs.~\eqref{eq:continuum-equations} highlights the importance of the turbulence effects on the average behaviour of the particle phase, and for the case of the linear drag law in Eqs.~\eqref{eq:linear-drag} these quantities are given by \cite{Swailes1997}
\begin{subequations}
    \label{eq:dispersion-coefficients}
    \begin{align}
        \overline{\boldsymbol{\kappa}} (\bm{r},t) & = \beta \int_{t_0}^t \left\langle \boldsymbol{\mathcal{H}}^{\top} [t;t^{\prime}] : \nabla \bm{R} (\bm{r}_p^{\prime},t^{\prime}; \bm{r},t) \right\rangle_{\bm{r}} \, dt^{\prime} \, ,\label{eq:kappa} \\
        \overline{\boldsymbol{\lambda}} (\bm{r},t) & = \beta \int_{t_0}^t \left\langle \boldsymbol{\mathcal{H}} [t;t^{\prime}] \cdot \bm{R} (\bm{r}_p^{\prime},t^{\prime}; \bm{r},t) \right\rangle_{\bm{r}} \, dt^{\prime} \, ,\label{eq:lambda} \\
        \overline{\boldsymbol{\mu}} (\bm{r},t) & = \beta \int_{t_0}^t \left\langle \dot{\boldsymbol{\mathcal{H}}} [t;t^{\prime}] \cdot \bm{R} (\bm{r}_p^{\prime},t^{\prime}; \bm{r},t) \right\rangle_{\bm{r}} \, dt^{\prime} \, , \label{eq:mu}
    \end{align}
\end{subequations}
in which $\bm{r}_p^{\prime} = \bm{r}_p(t^{\prime})$, and $\langle \cdot \rangle_{\bm{r}}$ denotes a conditional ensemble average involving only the subset of trajectories $\bm{r}_p$ such that $\bm{r}_p(t) = \bm{r}$. In Eqs.~\eqref{eq:dispersion-coefficients}, $\bm{R}$ is the Eulerian two-point, two-time correlation tensor of the fluctuating effective relative fluid velocity
\begin{equation} \label{eq:correlation-tensor}
    \bm{R} (\bm{r}^{\prime},t^{\prime}; \bm{r},t) = \langle \bm{u}^{\prime}(\bm{r}^{\prime},t^{\prime}) \bm{u}^{\prime}(\bm{r},t) \rangle \, ,
\end{equation}
and $\boldsymbol{\mathcal{H}}$ is the particle response tensor which, for the linear drag law \eqref{eq:linear-drag}, satisfies \cite{Bragg2012}
\begin{equation} \label{eq:response-tensor}
    \ddot{\boldsymbol{\mathcal{H}}} = -\beta \dot{\boldsymbol{\mathcal{H}}} + \beta \nabla \bm{u}(\bm{r}_p,t) \cdot \boldsymbol{\mathcal{H}} \, ,
    \qquad
    \begin{array}{ll}
        \boldsymbol{\mathcal{H}} [t^{\prime};t^{\prime}] & \hspace{-2mm} = \bm{0} \\
        \dot{\boldsymbol{\mathcal{H}}} [t^{\prime};t^{\prime}] & \hspace{-2mm} = \beta \mathbf{I}
    \end{array}
    \, .
\end{equation}

{\color{corr} The kinetic dispersion tensor coefficients $\overline{\boldsymbol{\kappa}}, \overline{\boldsymbol{\lambda}}, \overline{\boldsymbol{\mu}}$ in Eqs.~\eqref{eq:dispersion-coefficients} are history integrals that capture the influence of the flow field on particle trajectories through the action of the response tensor $\boldsymbol{\mathcal{H}}$. More specifically, $\overline{\boldsymbol{\kappa}}$ describes spatial convection and behaves as a momentum source, $\overline{\boldsymbol{\lambda}}$ describes spatial diffusion and behaves as a diffusion source, and $\overline{\boldsymbol{\mu}}$ describes velocity diffusion of particles and behaves as a stress source \cite{Skartlien2009}. In this sense, $\overline{\boldsymbol{\mu}}$ accounts for the multivaluedness of inertial particles in velocity space, and captures the effect of the turbulence so that clustering mechanisms such as caustics are included in the description provided by the kinetic PDF approach.
	
In essence, the particle response tensor $\boldsymbol{\mathcal{H}}$ describes the effect of a perturbation in the fluid velocity on the particle trajectory at subsequent times. The appearance of the carrier flow velocity gradient sampled by particles in Eq.~\eqref{eq:response-tensor} ensures that $\boldsymbol{\mathcal{H}}$ captures the dependence of particle trajectories on their path history through the flow field. Consequently, $\boldsymbol{\mathcal{H}}$ captures the behaviour of clustering mechanisms where $\nabla \bm{u}$ plays a central role, such as centrifuging and the non-local path history effect. In tandem, the correlation tensor $\bm{R}$ provides the description of the spatial structure of the carrier flow that is needed to account for clustering mechanisms where this is important, such as centrifuging and the sweep-stick mechanism.

It is therefore evident that both the path history described by $\boldsymbol{\mathcal{H}}$ and the spatiotemporal description of the flow field provided by $\bm{R}$ are necessary to describe the range of clustering mechanisms outlined in Section \ref{sec:mechanisms}. This is embodied in the mathematical form of the kinetic dispersion tensors in Eqs.~\eqref{eq:dispersion-coefficients}, with it being the correlation between $\boldsymbol{\mathcal{H}}$ and $\bm{R}$ that accounts for the effect of turbulence on particles within the kinetic dispersion tensors.} The emergence of this correlation within history integrals in Eqs.~\eqref{eq:dispersion-coefficients} demonstrates that these dispersion tensors are hysteretic in nature, and highly dependent on the path history of particles as they move through the carrier flow field. The resultant description provided by the dispersion tensors is therefore central to representing the detailed behaviour of particle dispersion in turbulence, and in particular that observed in clustering.

\subsection{The particle momentum equation interpreted as a convection-diffusion balance} \label{sec:mass-flux}

Of particular interest in the context of clustering as viewed from a particle pair description is the relative contribution of the different fluxes to the particle phase momentum balance. To elucidate the significance of the dispersion coefficients \eqref{eq:dispersion-coefficients} within these contributions, it is helpful to recast the momentum equation \eqref{eq:momentum-equation} in terms of the particle mass flux $\rho \overline{\bm{w}}$. Using Eq.~\eqref{eq:linear-drag} this can be written in the convective-diffusive form
\begin{widetext}
	\begin{equation} \label{eq:mass-flux}
		\rho \overline{\bm{w}}
		= \overbrace{ \rho \Bigg[
			\left\langle \bm{u} \right\rangle
			+ \tau_p \Bigg\{
			\underbrace{ \left[ \overline{\boldsymbol{\kappa}} - \nabla \cdot \overline{\boldsymbol{\lambda}} \right] }_{\text{\normalsize $\bm{d}^1$}}
			\underbrace{ - \nabla \cdot \overline{\bm{c}\bm{c}} }_{\text{\normalsize $\bm{d}^2$}}
			\underbrace{ - \frac{D \overline{\bm{w}}}{D t} }_{\text{\normalsize $\bm{d}^3$}} \Bigg\} \Bigg] }^{\text{{\normalsize Convective flux}}}
		- \overbrace{ \underbrace{ \tau_p \left( \overline{\bm{c}\bm{c}} + \overline{\boldsymbol{\lambda}}^{\top} \right)}_{\text{\normalsize $\bm{D}$}} \cdot \nabla \rho }^{\text{\normalsize Diffusive flux} }
		\, .
	\end{equation}
\end{widetext}

In Eq.~\eqref{eq:mass-flux}, two fundamental contributions exist to the net forces acting on an elemental volume of particles at equilibrium. The first of these is a kinetic contribution arising from the particle kinetic stresses $\overline{\bm{c}\bm{c}}$, and appears within the convective flux as the turbophoretic term $\bm{d}^{2}$, given by the divergence of the kinetic stresses, and also as a contribution to the particle dispersion via the particle diffusion coefficient $\bm{D}$. The second is a contribution from the net turbulent aerodynamic force acting on the particles, which itself is composed of two contributions that emanate from the kinetic dispersion tensors. Firstly is the spatial gradient of a turbulent stress $\overline{\boldsymbol{\lambda}} \rho$ which has two components \cite{Reeks1991}; a convective contribution $\nabla \cdot \overline{\boldsymbol{\lambda}}$ that has the same effect as the turbophoresis, and a diffusive contribution $\overline{\boldsymbol{\lambda}}^{\top}$ that appears within $\bm{D}$. Secondly, a body force $\overline{\boldsymbol{\kappa}}$ arises as a result of trajectory biasing of the particle motion in a particular direction, either as a result of the imposition of an external force like gravity \cite{Stafford2021}, or as a result of inhomogeneity in the turbulence of the suspending carrier flow, such as in a turbulent boundary layer \cite{Bragg2012a}. The manner in which these contributions act is distinct; in contrast to the clear interpretation of $\overline{\boldsymbol{\kappa}}$ as a body force, the contributions from $\nabla \cdot \overline{\boldsymbol{\lambda}}$ and the turbophoretic term $\bm{d}^{2}$ act together as a stress-induced phoresis that is responsible for the migration of particles in the direction of decreasing turbulent stresses \cite{Brenner2011}.

Both the kinetic and the turbulent stress components make important contributions to the force balance \eqref{eq:mass-flux} that acts on an elemental volume of particles at equilibrium. This is manifest as both a diffusive contribution that is dependent on the particle concentration gradient, and a convective contribution that is proportional to the mean particle concentration. The diffusive flux $\bm{D}$ is seen to depend on both the kinetic stresses $\overline{\bm{c}\bm{c}}$
and the spatial diffusion of particles described by $\overline{\boldsymbol{\lambda}}$. In the convective flux the three key terms $\bm{d}^{1}$, $\bm{d}^{2}$, and $\bm{d}^{3}$ emerge (in addition to the mean fluid velocity $\langle \bm{u} \rangle$).
The inertial term $\bm{d}^{3}$ clearly vanishes for systems which can be considered both statistically stationary and homogeneous, and it will also not contribute when, as will be the case here, $\overline{\bm{w}} = \bm{0}$.
The turbophoretic contribution $\bm{d}^{2}$ has been accounted for in some existing particle pair models \cite{Zaichik2007,CHUN2005}.
The contribution from  term $\bm{d}^{1}$ reflects the non-local turbulent drift experienced by particles as an accumulation of trajectory history effects that act directly as a momentum source, and is captured through the dispersion tensors of the kinetic model.
Specifically, $\bm{d}^{1}$ combines the effect of the turbulent stresses described by $\overline{\boldsymbol{\lambda}}$ and the effect of body forces in $\overline{\boldsymbol{\kappa}}$, and together these form a contribution that is hysteretic in nature, in contrast to the kinetic form of $\bm{d}^{2}$.
Within the context of a steady-state particle pair framework the total mass flux $\rho \overline{\bm{w}}$ is zero, and it is therefore the balance of convective to diffusive flux contributions that determines the equilibrium distribution of $\rho$, and thereby the RDF that characterises the degree of clustering experienced by particles.

It should be noted that the various flux contributions that appear in Eq.~\eqref{eq:mass-flux} are a direct consequence
of a kinetic approach that accounts explicitly for the spatiotemporal correlation structure of the flow via the correlation tensor
$\bm{R}$ in the dispersion tensors, Eq.~\eqref{eq:dispersion-coefficients}.
In the absence of this level of description in the modelling, with approximate white-in-time correlations introduced,
the kinetic equation, Eq.~\eqref{eq:kinetic-equation}, reduces to the Fokker-Planck equation.
Whilst still capable of describing the systemic behaviour of particles with high inertia, in this case the dispersion tensors
become constant, and the non-local turbulent drift $\bm{d}^{1}$
vanishes completely, leaving the turbophoresis $\bm{d}^{2}$
as the sole convective contribution to the particle mass flux balance.
The Fokker-Planck equation is itself a generalisation which corresponds to a Wiener process with a drift term that accounts for the drag force on a particle, without which the particle dynamics reduce to Brownian motion and are governed by the Stokes-Einstein relation \cite{Reeks2021}. The particle behaviour is purely diffusive in this case, with the significance of the kinetic equation \eqref{eq:kinetic-equation} then being clear that it extends the classical work on the kinetic theory of gases to include the effects of turbulence on inertial particles through both the appearance of the convective flux contributions $\bm{d}^{1}$ and $\bm{d}^{2}$,
and also the specification of an expression for the effective diffusion coefficient $\bm{D}$ in Eq.~\eqref{eq:mass-flux}.

\subsection{Analysis of existing particle pair models}

Of particular relevance in the consideration of the mass flux balance equation \eqref{eq:mass-flux} is that previous modelling approaches for dispersed particle transport have neglected either one or both terms in the non-local turbulent drift $\bm{d}^{1}$.
In the case of one-particle PDF models this includes the assumption that $\overline{\boldsymbol{\kappa}} \equiv \bm{0}$ in arbitrary homogeneous turbulent flows \cite{Reeks1992,Elghobashi1994}, whilst for particle pair models representations have been proposed which either do not explicitly take into account the contribution of
$\bm{d}^{1}$ \
\cite{CHUN2005},
or assume that it is identically zero across all values of the particle Stokes number $St$ \cite{Zaichik2007}.
However, it has been formally demonstrated that
$\bm{d}^{1} \equiv \bm{0}$
only in the limit of fluid tracers ($St = 0$), a result known as the \emph{fully mixed condition} \cite{Bragg2012}, and there is no obvious reason why this should be the case in the general situation of inertial particles.

In the case of the model developed by Chun \textit{et al}. \cite{CHUN2005}, the use of a perturbation expansion to develop a model description applicable for $St \ll 1$ has the consequence that the PDF equation governing the distribution of the pair separation includes neither of the convective fluxes associated with non-local turbulent drift $\bm{d}^{1}$ or turbophoresis $\bm{d}^{2}$
in Eq.~\eqref{eq:mass-flux} explicitly. This is due to these contributions being embedded within the higher-order terms which are omitted during truncation of the expansion to $\mathcal{O}(St)$. The form of the drift flux term which emerges from the modelling procedure is a leading-order approximation to $\bm{d}^{1}$, which despite it maintaining a non-local description, does not capture the path history mechanism embodied in $\bm{d}^{1}$, as identified in the analysis of Bragg \& Collins \cite{Bragg2014a}. The PDF equation in this case does not contain a term with the form of $\bm{d}^{2}$,
and this contribution to clustering is therefore omitted entirely in the model description.

Whilst the work of Zaichik \& Alipchenkov \cite{Zaichik2003,Zaichik2007,Zaichik2009} uses a kinetic description to statistically analyse the separation of inertial particle pairs, one shortcoming of the modelling procedure is that the non-local turbulent drift contribution
$\bm{d}^{1}$
is identically zero. However, this emerges only as a consequence of the particular choice of closure models that are employed, rather than being an explicit independent assumption.
Specifically, within the dispersion tensors given in Eqs.~\eqref{eq:dispersion-coefficients} the particle response tensor
$\boldsymbol{\mathcal{H}}$ is approximated by the Green's function for the  linear drag law, Eq.~\eqref{eq:linear-drag},
yielding a trajectory independent description.
Similarly, the fluid correlation tensor $\bm{R}$, which for the velocity field Eq.~\eqref{eq:inhomogeneous-flu-vel} describes the Lagrangian two-point structure function in homogeneous turbulence, is closed at a local level, again leading to a trajectory independent representation. Consequently, this results in the models for both $\boldsymbol{\mathcal{H}}$ and $\bm{R}$ being completely deterministic, with the resultant form of the dispersion tensors Eqs.~\eqref{eq:dispersion-coefficients} leading to the approximation
$\bm{d}^{1} = \bm{0}$.
To achieve a closure where this is not the case, it is necessary to formulate a non-local model which is able to account for the path history effects along trajectories \cite{Bragg2012a,Stafford2021}. It is worth noting however that the model used for the Lagrangian two-point structure function \cite{Zaichik2007} contains an additional transport term that accounts for turbulence unsteadiness and transport of fluid velocity fluctuations along a particle pair trajectory. It is this level of description which is able to capture the turbophoretic contribution to the particle behaviour, and thereby the enhancement in clustering which is observed.

More recent work on particle pair modelling \cite{Belan2014}, employing an analysis based upon a Fokker-Planck equation for the particle PDF, found that the net particle mass flux in the direction opposite to the turbophoresis only exists for sufficiently inertial particles. However, this conclusion stems from the approximation that the fluid velocity is delta-correlated in time, which captures well the dynamics of sufficiently inertial particles, but remains a poorer descriptor of low-inertia particles \cite{Reeks2021}. Another study has proposed a model which states that particles only ever migrate towards the wall within the logarithmic region of the boundary layer \cite{Sikovsky2019}, which does not account for the possibility of term $\bm{d}^{1}$ in Eq.~\eqref{eq:mass-flux} being significant.

\subsection{Role of the Current Work}

In light of the omissions made in the modelling procedures of previous studies, the purpose of the present work is to demonstrate that in the context of the clustering of particles the non-local turbulent drift $\bm{d}^{1}$ in
Eq.~\eqref{eq:mass-flux} is non-zero, and furthermore can become the dominant contribution to the convective flux in some flow configurations. Such a demonstration is with a view to providing verification of important features associated with the specific form of kinetic approach
outlined in Section \ref{sec:PDF-2p}, and in particular how the role of turbulence is captured \cite{Reeks2021}. It is worth highlighting that this form of kinetic model provides a description which is free of spurious drift \cite{Bragg2012}, and is therefore suitable for
detailed analysis of the various contributions to the particle mass flux balance equation \eqref{eq:mass-flux}.
Further, the ability of the kinetic PDF approach to capture the subtle effects of turbulence on particle behaviour across the entire range of $St$ has previously been examined in the context of gravitational settling \cite{Stafford2021}, where the net settling velocity of particles is modified due to preferential sampling of the turbulent flow structures by particles. The present work serves to illustrate that the kinetic PDF approach is also capable of providing such a description in the case of inhomogeneous turbulence.

\section{Inhomogeneous Flow Field Model} \label{sec:inhomogeneous-flow-model}

\subsection{Properties}

Since the form of $\bm{u}(\bm{r},t)$ specified in Eq.~\eqref{eq:inhomogeneous-flu-vel} is determined directly from the homogeneous velocity field $\bm{U}(\bm{r},t)$, relevant properties of $\bm{u}(\bm{r},t)$ can be deduced in a straightforward manner from those of $\bm{U}(\bm{r},t)$.
Most notably it is clear that
$\bm{u}(\bm{0},t) = \bm{0}$, so that $\bm{r} = \bm{0}$ can be interpreted as a stagnation point.
Additionally it is straightforward to show from Eq.~\eqref{eq:inhomogeneous-flu-vel} that if $\bm{U}(\bm{r},t)$ is defined as an incompressible, periodic, and zero-mean flow field, then  $\bm{u}(\bm{r},t)$ must also satisfy all these properties.

An important consequence of using Eq.~\eqref{eq:inhomogeneous-flu-vel} to define $\bm{u}(\bm{r},t)$ is that the fluctuating fluid velocity correlations $\bm{R}$ are then expressible in terms of the correlations of the underlying homogeneous velocity field $\bm{U}(\bm{r},t)$
\begin{align} \label{eq:two-point-correlation}
    \bm{R} (\bm{r}^{\prime},t^{\prime}; \bm{r},t) & = \big\langle \bm{U}^{\prime} (\bm{r}^{\prime},t^{\prime}) \bm{U}^{\prime} (\bm{r},t) \big\rangle - \big\langle \bm{U}^{\prime} (\bm{r}^{\prime},t^{\prime}) \bm{U}^{\prime} (\bm{0},t) \big\rangle \nonumber \\
    & - \big\langle \bm{U}^{\prime} (\bm{0},t^{\prime}) \bm{U}^{\prime} (\bm{r},t) \big\rangle + \big\langle \bm{U}^{\prime} (\bm{0},t^{\prime}) \bm{U}^{\prime} (\bm{0},t) \big\rangle
    \, .
\end{align}
Therefore, if the underlying velocity field $\bm{U}(\bm{r},t)$ is isotropic
then two-point two-time correlation tenor $\bm{R}$ for the effective relative velocity field $\bm{u}(\bm{r},t)$, given by Eq.~\eqref{eq:inhomogeneous-flu-vel}, can be represented as a superposition of isotropic correlation tensors, even though this velocity field is inhomogeneous and anisotropic.
Writing $\bm{Q} (\bm{r} - \bm{r}^{\prime}) = \big\langle \bm{U}^{\prime} (\bm{r}^{\prime},t) \bm{U}^{\prime} (\bm{r},t) \big\rangle$, the correlations appearing in Eq.~\eqref{eq:two-point-correlation} can therefore all be written as functions of a single spatial argument, which for a one-time correlation gives
\begin{equation} \label{eq:inhomogeneous-correlations-decomp}
    \bm{R} (\bm{r}^{\prime},t; \bm{r},t) = \bm{Q} (\bm{r} - \bm{r}^{\prime}) - \bm{Q} (\bm{r}^{\prime}) - \bm{Q} (\bm{r}) + \bm{Q} (\bm{0})
    \, .
\end{equation}
Further, from Eq.~\eqref{eq:inhomogeneous-correlations-decomp} the one-point one-time correlations of
$\bm{u}(\bm{r},t)$ are given by
\begin{equation} \label{eq:inhomogeneous-reynolds-stresses}
    \bm{R} (\bm{r},t; \bm{r},t) = 2 \big[ \bm{Q} (\bm{0}) - \bm{Q} (\bm{r}) \big]
    \, .
\end{equation}
Consequently the mean square fluid velocities of the flow field $\bm{u}(\bm{r},t)$ are spatially dependent,
which exemplifies the desired inhomogeneity of this configuration.

\subsection{Representation in spherical-polar coordinates}

\begin{figure}
    \includegraphics[width=\columnwidth,trim={0 0 0 0}]{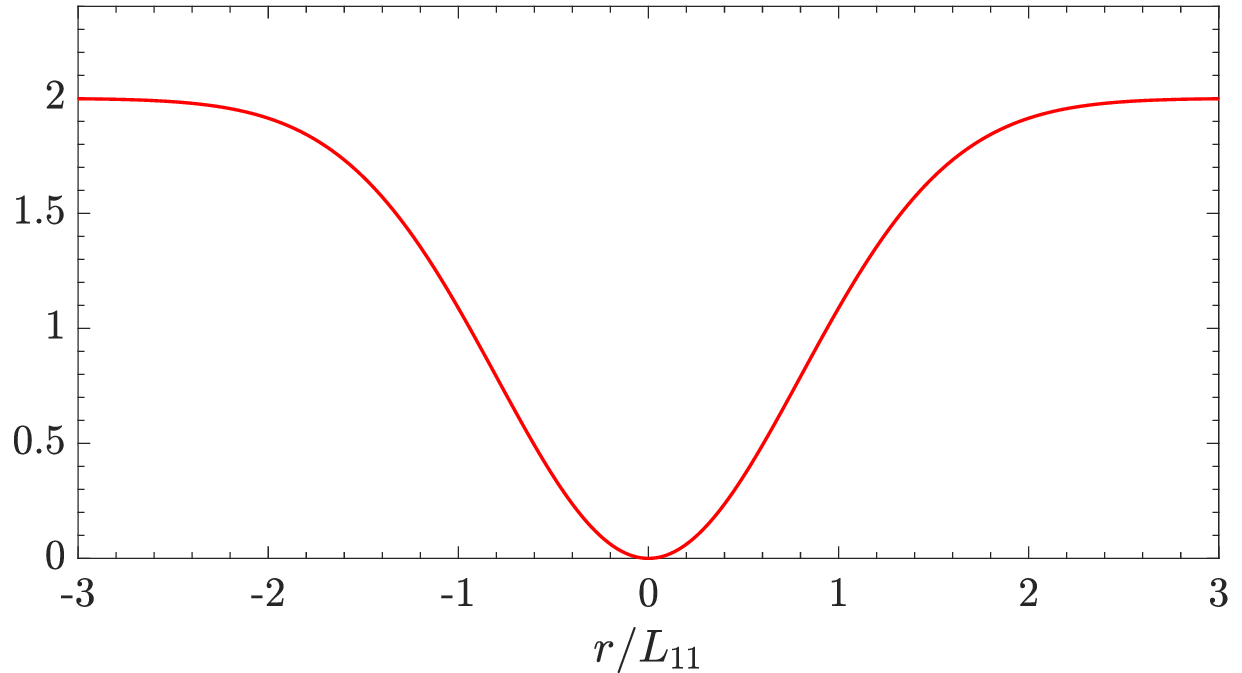}
    \caption{Longitudinal mean-square velocity $R_{\parallel} = \langle \bm{u}^{\prime} \bm{u}^{\prime} \rangle_{\parallel}$}
    \label{fig:RMSI_plot}
\end{figure}

The flow field defined by Eq.~\eqref{eq:inhomogeneous-flu-vel} describes fluid velocities relative to that
experienced by the target particle identified with $\bm{r} = \bm{0}$.
The statistics associated with this flow and, in consequence, the resulting long-time (statistically steady)
particle dynamics will exhibit spherical symmetry about this point.
This makes it natural to represent the various tensor quantities in terms of their
components with respect to the standard spherical co-ordinate basis
$\{ \bm{e}_{r} , \bm{e}_{\theta}, \bm{e}_{\phi} \}$.
For example, with respect to the standard Cartesian basis $\{ \bm{e}_{i} \}$
the isotropic correlation tensor $\bm{Q}$ has representation
$\bm{Q} = Q_{ij} \bm{e}_{i} \bm{e}_{j}$ with\cite{batchelor1953theory}
\begin{equation} \label{eq:isotropic-correlation-tensor}
    Q_{ij} (\bm{r}) = {u^{\prime}}^{2} \left[ g(r) \delta_{ij} + \left( f(r) - g(r) \right) \frac{r_i r_j}{r^2} \right]
    \, ,
\end{equation}
where $r = \lvert \bm{r} \rvert$,
$u^{\prime}$ the root mean square fluid velocity,
and $f$, $g$ the normalised longitudinal and lateral correlation coefficients associated with $\bm{U}$.
With respect to the spherical co-ordinate basis this simplifies to 
$
\bm{Q} =
Q_{\parallel} \bm{e}_{r} \bm{e}_{r} +
Q_{\perp}
(
\bm{e}_{\theta} \bm{e}_{\theta} + \bm{e}_{\phi} \bm{e}_{\phi}
)
$
with longitudinal and lateral components depending only on the separation distance $r$,
\begin{equation}
Q_{\parallel} = {u^{\prime}}^2 f ( r ) \ , 
\quad
Q_{\perp} = {u^{\prime}}^2 g ( r ) \ .
\end{equation}
Correspondingly, the one-point correlations for the inhomogeneous flow field $\bm{u} (\bm{r},t)$, as given by Eq.~\eqref{eq:inhomogeneous-reynolds-stresses}, have the components
\begin{equation}
R_{\parallel} = 2 {u^{\prime}}^2 \bigl ( 1 - f ( r ) \bigr ) \ , 
\quad
R_{\perp} = 2 {u^{\prime}}^2 \bigl ( 1 - g ( r ) \bigr ) \ .
\end{equation}
Thus, these one-point correlations of $\bm{u} (\bm{r},t)$ depend only upon the radial displacement $r$ from the origin.
For the choice of longitudinal correlation function $f(r)$ specified in Eq.~\eqref{eq:longitudinal-correlation-coefficient},
the variation in the mean square fluid velocity profile is shown in Fig.~\ref{fig:RMSI_plot},
where the radial displacement has been normalized using the turbulent integral length scale $L_{11}$.
At large radial displacements the correlation profile assumes a constant value associated with an essentially homogeneous field.
However at small displacements the correlation decrease toward zero producing a region of marked inhomogeneity, with
$\langle \bm{u}^{\prime} \bm{u}^{\prime} \rangle = \bm{0}$ at the origin.
In a similar way the dispersion tensors that determine the particle pair PDF have simplified
representations
\begin{subequations}
\begin{align}
{\bm{\kappa}} & =
{\kappa}_{\parallel} \bm{e}_{r}
\\
{\bm{\lambda}} & =
{\lambda}_{\parallel} \bm{e}_{r} \bm{e}_{r} +
{\lambda}_{\perp}
\big (
\bm{e}_{\theta} \bm{e}_{\theta} + \bm{e}_{\phi} \bm{e}_{\phi} 
\big )
\end{align}
\end{subequations}
where spatial dependence in the the polar components ${\kappa}_{\parallel}$, ${\lambda}_{\parallel}$, ${\lambda}_{\perp}$ is solely in terms of the separation $r$.
It follows that the associated moments of the PDF $p$ will also depend only on this spatial separation.
In particular the representation of the RDF can be expressed in a simple form.

\subsection{An expression for the radial distribution function}

In a statistically steady-state $\rho \overline{\bm{w}} = \bm{0}$, and the radial component of 
Eq.~\eqref{eq:mass-flux} reduces to
\begin{equation}
\label{eq:zero-flux}
0 = \rho d_{\parallel} - D_{\parallel} \dfrac{d \rho}{d r}
\end{equation}
giving an expression for the radial distribution function (RDF)
\begin{equation} \label{eq:RDF}
\rho (r) = \rho_{\infty} 
\exp
\left[
- \int_{r}^{\infty} \frac{d_{\parallel}}{D_{\parallel}} \, dr^{\prime}
\right]
\end{equation}
where $\rho_{\infty} = \lim\limits_{r \rightarrow \infty} \rho (r)$ is a specified uniform particle concentration at large separations, and the radial drift and diffusion coefficients, $d_{\parallel}$ and $D_{\parallel}$,  are obtained from Eq.~\eqref{eq:mass-flux}.
Specifically, since $\langle \bm{u} \rangle = \overline{\bm{w}} = \bm{0}$,
\begin{subequations}
\label{eq:d-D-radial}
\begin{align}
d_{\parallel}
& =
\tau_p
\left (
\overline{\kappa}_{\parallel} - \left ( \nabla \cdot \overline{\bm{\lambda}} \right )_{\parallel} -
\left ( \nabla \cdot \overline{\bm{cc}} \right )_{\parallel}
\right ) 
= d^{1}_{\parallel} + d^{2}_{\parallel} \ ,
\label{eq:d-radial}
\\
D_{\parallel} 
& = 
\tau_p
\left (
\overline{\bm{cc}}_{\parallel} + \overline{\lambda}_{\parallel}
\right ) \ .
\label{eq:D-radial}
\end{align}
\end{subequations}
Note that these expressions involve both the longitudinal and lateral components of 
$\overline{\bm{\lambda}}$ and $\overline{\bm{cc}}$;
\begin{equation}\label{eq:div-lambda}
\bigl (
\nabla \cdot \overline{\bm{\lambda}}
\bigr )_{\parallel}
=
\dfrac{n-1}{r}
\bigl (
\overline{\lambda}_{\parallel} - \overline{\lambda}_{\perp}
\bigr ) 
+
\dfrac{d}{dr} \overline{\lambda}_{\parallel}
\end{equation}
and similarly for $( \nabla \cdot \overline{\bm{cc}} )_{\parallel}$.
In Eq.~\eqref{eq:div-lambda} $n=3$ or $n=2$; for 3D  or 2D polar co-ordinates, respectively.

Since the RDF given by Eq.~\eqref{eq:RDF} quantifies the degree of particle clustering in the system,
the influence of the dispersion coefficients
$\overline{\bm{\kappa}}$, $\overline{\bm{\lambda}}$ becomes apparent,
and the relative contributions of the various terms in Eqs.~\eqref{eq:d-D-radial} are of significance.
As already discussed, other studies\cite{CHUN2005,Zaichik2007} have developed expressions for the RDF that do not
involve all of the contributions that emerge from the kinetic approach used here.
In particular, the contribution from the non-local drift flux 
$\bm{d}^{1}$ is absent. It is therefore important
to assess the role and significance of this term in the mechanisms that act to increase
particle clustering.
Section \ref{sec:numerics} presents results obtained from exact computations of the terms in 
Eqs.~\eqref{eq:d-D-radial}, via the underlying definitions given by Eqs.~\eqref{eq:dispersion-coefficients},
and also compares corresponding profiles of the RDF, Eq.~\eqref{eq:RDF}, with those
obtained directly from particle trajectory simulations. Prior to that, in the next Section, approximations to the
dispersions tensors $\overline{\bm{\kappa}}$, $\overline{\bm{\lambda}}$, $\overline{\bm{\mu}}$ are
developed. These are based on assumptions appropriate for homogeneous systems, and
therefore serve as references against which the influence of the imposed inhomogeneities can be assessed.

\section{Local homogeneous approximation of the PDF dispersion tensors} \label{sec:LHA}

The starting point for modelling the dispersion tensors \eqref{eq:dispersion-coefficients} is the construction of simple approximations
that are local to the point $\bm{r}$ at which the description is considered.
The approach assumes a quasi-homogeneous flow and particle distribution
in the vicinity of $\bm{r}$ and, consequently, is referred to as a \textit{local homogeneous approximation} (LHA).
These assumptions are applied to the particle response tensor $\boldsymbol{\mathcal{H}}$ and correlation tensor
$\bm{R}$, defined in Eqs.~\eqref{eq:correlation-tensor}, \eqref{eq:response-tensor},
and which together determine the unclosed conditional average within the dispersion tensors
Eq.~\eqref{eq:dispersion-coefficients}.

In the case of $\bm{R}$, a locality assumption is applied by supposing that $\bm{r}_{p}^{\prime} \approx \bm{r}$, from which decomposition into a product of one-point and two-time correlations follows \cite{Swailes1999}
\begin{equation} \label{eq:LHA-correlation-tensor}
\bm{R} ( \bm{r}_{p}^{\prime},t^{\prime}; \bm{r},t ) \approx \bm{R} ( \bm{r},t; \bm{r},t ) \, E (t - t^{\prime}) \, ,
\end{equation}
in which $E$ is introduced as a model for the temporal decorrelation of the velocity field $\bm{u}(\bm{r},t)$ along particle trajectories.
The Reynolds stresses associated with $\bm{u} (\bm{r},t)$ are then specified as in
Eq.~\eqref{eq:inhomogeneous-reynolds-stresses},
whilst an exponential temporal decorrelation is taken as
\begin{equation} \label{eq:LHA-temporal-correlation}
E (t - t^{\prime}) = \exp \left[ - \frac{1}{\tau_{Lp}} (t - t^{\prime}) \right] \, ,
\end{equation}
Here $\tau_{Lp}$ is the Lagrangian fluid integral timescale for the particle-sampled fluid
velocity $\bm{u} (\bm{r}_{p}^{\prime},t^{\prime})$.
A suitable expression for $\tau_{Lp}$ in homogeneous flows,
as obtained from the curve fitting of DNS data\cite{Wang1993}, is given by
\begin{equation} \label{eq:lagrangian-integral-timescale}
\tau_{Lp} = \tau_{E} - \frac{\tau_{E} - \tau_{L}}{\left( 1 + St_{E} \right)^{-0.4 \left( 1 + 0.01 \, St_{E} \right)}} \, ,
\end{equation}
where $\tau_{E}$ is the Eulerian fluid integral timescale, $\tau_{L}$ is the Lagrangian fluid integral timescale along fluid element trajectories, and $St_{E}$ is the Stokes number defined in terms of the Eulerian integral timescale,
\textit{i.e.}~$St_{E} = \tau_{p} / \tau_{E}$.
The model for $\bm{R}$, Eq.~\eqref{eq:LHA-correlation-tensor}, therefore accounts for the effects of particle inertia through the single quantity $\tau_{Lp}$.

For the particle response tensor $\boldsymbol{\mathcal{H}}$, a corresponding approximation is made by neglecting the contribution from the fluid velocity gradient within the governing equation for $\boldsymbol{\mathcal{H}}$, Eq.~\eqref{eq:response-tensor}.
This has the effect of making $\boldsymbol{\mathcal{H}}$ trajectory independent, and the resulting simplified governing equation yields the approximation \cite{Swailes1999}
\begin{equation} \label{eq:greens-tensor}
\mathbf{H} [t;t^{\prime}] = \big( 1 - \exp \left[ -\beta (t - t^{\prime}) \right] \big) \mathbf{I} \, .
\end{equation}
The approximations Eq.~\eqref{eq:LHA-correlation-tensor} for $\bm{R}$
and Eq.~\eqref{eq:greens-tensor} for $\boldsymbol{\mathcal{H}}$
are both deterministic quantities, meaning that evaluation of the dispersion tensors
$\overline{\boldsymbol{\kappa}}$,
$\overline{\boldsymbol{\lambda}}$, and $\overline{\boldsymbol{\mu}}$ using these expressions no longer requires closure of the conditional averages which appear within Eqs.~\eqref{eq:dispersion-coefficients}.
Subsequent evaluation of the integrals and taking the limit $t \to \infty$ yields the steady-state LHA for the dispersion tensors,
which in polar components for the inhomogeneous flow field \eqref{eq:inhomogeneous-flu-vel} are
\begin{equation}
    \label{eq:LHA-dispersion-coefficients}
    \begin{array}{ll}
        \overline{\kappa}^{\text{L}}_{\parallel}  = {0} \, , & \\[1em]
        \overline{\lambda}^{\text{L}}_{\parallel}  = \sigma^2 \bigl ( 1 - f ( r )  \bigr )
        \, , &
 \overline{\mu}^{\text{L}}_{\parallel}  = \tau_{Lp}^{-1} \,  \overline{\lambda}^{\text{L}}_{\parallel} \, , \\[1em]
      \overline{\lambda}^{\text{L}}_{\perp} =  \sigma^2  \bigl ( 1 - g ( r ) \bigr )
        \, , &
        \overline{\mu}^{\text{L}}_{\perp} = \tau_{Lp}^{-1}  \,\overline{\lambda}^{\text{L}}_{\perp} \, ,
    \end{array}
\end{equation}
where $\sigma^2 =  2 {u^\prime}^2 / \bigl ( St_{Lp} ( 1 + St_{Lp} ) \bigr )$
with
$St_{Lp} = \tau_{p} / \tau_{Lp}$ being the associated Stokes number.

The approximations given by Eqs.~\eqref{eq:LHA-dispersion-coefficients} represent
the simplest closures for the dispersion tensors in the flow field $\bm{u}$,
with the assumption of locality meaning that trajectory history effects are neglected at this level of description.
The result $\overline{\kappa}^{\text{L}}_{\parallel} = 0$ follows from the isotropy of the approximation
Eq.~\eqref{eq:greens-tensor} for $\boldsymbol{\mathcal{H}}$ which, in turn, means that
$\nabla \bm{R}$ in Eq.~\eqref{eq:kappa} contracts to zero in view of the assumed incompressibility of $\bm{u}$.
Further, for the same reason, the relation between $f(r)$ and $g(r)$ implies that
$\nabla \cdot \overline{\bm{\lambda}}^{\text{L}} = \bm{0}$.
As a result the LHA fails to capture any of the non-local turbulent drift flux described by term
$\bm{d}^{1}$ in Eq.~\eqref{eq:mass-flux}.

With regard to the turbophoretic contribution $\bm{d}^{2}$ in Eq.~\eqref{eq:mass-flux}, specification of the kinetic stresses
$\overline{\bm{c}\bm{c}}$ formally requires solution of the kinetic stress transport equation
Eq.~\eqref{eq:kinetic-stress-equation}, which itself requires further closures to be made, and does not yield analytical solutions \cite{Zaichik2003}.
Notwithstanding this, it is possible to construct a simple approximation for the kinetic stresses within the framework of the kinetic approach by using assumptions that are consistent with the LHA.
Specifically, taking the steady state form of Eq.~\eqref{eq:kinetic-stress-equation} in the case of the linear drag law given by 
Eq.~\eqref{eq:linear-drag}, neglecting spatial gradients, and making the approximation
$\overline{\bm{c} \boldsymbol{\kappa}} \approx \bm{0}$ yields the model
\begin{equation} \label{eq:cc-LHA}
    \overline{\bm{c}\bm{c}}^{\text{L}} = \tau_{p} \, \overline{\boldsymbol{\mu}}^{\text{L}}  \, .
\end{equation}
Incompressibility of $\bm{u}$ then
implies $\nabla \cdot \overline{\bm{c}\bm{c}}^{\text{L}} = \bm{0}$
so that, as with $\bm{d}^{1}$, the turbophoretic contribution $\bm{d}^{2}$ is
not accounted for using these simple local homogeneous approximations.

Evaluation of RDF using these LHA models in Eq.~\eqref{eq:RDF},
produces the approximation $\rho (r) \approx \text{constant}$.
Therefore, modelling at the level of the LHA fails to retrieve any evidence of the expected build up in particle concentration that is characteristic of clustering within a particle pair framework. Notwithstanding this, the LHA does offer a simple model against which numerical results of the PDF dispersion tensors, and convective and diffusive mass flux contributions obtained
from particle tracking simulation data can be compared.

\section{Numerical Methodology} \label{sec:numerics}

The aim is to evaluate all of the terms in the convective and diffusive fluxes appearing in Eq.~\eqref{eq:mass-flux}
without recourse to any closure modelling.
To this end the dispersion tensors given by  Eqs.~\eqref{eq:dispersion-coefficients}
are computed exactly. This requires explicit evaluation of the underlying $\bm{r}$-conditioned
ensemble averages using simulations of particle trajectories within the inhomogeneous flow $\bm{u}$. The particle response tensor $\boldsymbol{\mathcal{H}}$ is evaluated along trajectories using Eq.~\eqref{eq:response-tensor}, whilst the Eulerian fluid correlation tensor $\bm{R}$ for the underlying flow field $\bm{U}$ is specified using the isotropic formulae Eq.~\eqref{eq:isotropic-correlation-tensor}, with the spatial separation $r$ between the particle position at times $t^{\prime}$ and $t$ then being used to evaluate $\bm{R}$ along trajectories. Computationally, it is convenient to evaluate the contributions from different trajectories to the dispersion tensors in Cartesian component form.
Conversion to spherical polar components is then easily affected using the standard change of basis matrix.

In this work the inhomogeneous velocity field $\bm{u}$
is generated using the technique of kinematic simulation (KS),
and the specification of this is outlined in Appendix \ref{sec:KS}.
The application of KS to studying particle-laden turbulent flows has been widely adopted due to its relatively low computational cost, and has produced insights into phenomena that have not been previously observed \cite{Maxey1987,Maxey1987a,IJZERMANS2010}. A particularly significant feature of KS
is the scope it affords to control flow structures,
so that the influence of spatial and temporal decorrelation on the resultant particle behaviour can be accurately quantified.
Furthermore, KS is able to generate velocity fields conforming to Gaussian distributions. This aligns
naturally with the kinetic framework, which provides an exact description of statistical dynamics for Gaussian fields.
It follows that KS offers a precise means of assessing the role of the associated dispersion tensors on clustering,
free from the influence of confounding factors. Also, importantly,
the expected behaviour of inertial particles is qualitatively reproduced by this approach, including the crossing trajectories effect, clustering, and pairwise dispersion \cite{Murray2016a}.

This work investigates the influence of particle inertia and turbulence intensity on the degree of clustering experienced by particles. To quantify the effect of variations in the turbulence intensity on the particle mass flux contributions, the notion of a \emph{turbulence structure parameter} is used \cite{Wang1993}. This is defined as the ratio between $u^{\prime}$ and the characteristic eddy velocity which is formed from the turbulent integral lengthscale $L_{11}$ and $\tau_{E}$
\begin{equation} \label{eq:turbulence-structure-parameter}
    \hat{u}^{\prime} = \frac{u^{\prime} \tau_{E}}{L_{11}} \, .
\end{equation}
The particle mass flux terms are analysed here for two values of each of the particle inertia ($St_E = 0.1, 1.0$) and the turbulence intensity ($\hat{u}^{\prime} = 1, 4$).

Particle trajectories determined by Eqs.~\eqref{eq:part-eq} and \eqref{eq:linear-drag}, and response tensor  by
Eq.~\eqref{eq:response-tensor} are numerically integrated using a fourth-order Adams predictor-corrector method, and an ensemble of $10^6$ particles is used to achieve statistical convergence.
For computational economy attention has been restricted to two-dimensional systems (so that $n=2$ in Eq.~\eqref{eq:div-lambda}).
Then, all particle phase statistics, including those for the dispersion tensors $\overline{\boldsymbol{\kappa}}$ and
$\overline{\boldsymbol{\lambda}}$, are evaluated conditionally on the radial coordinate $r$.
In particular, the steady state particle number density $\rho$ is calculated such that
\begin{equation} \label{eq:number-density-calculation}
    \rho (r) = \frac{N_r}{\delta A_r} \, ,
\end{equation}
where $N_r$ is the number of particles at radial displacement $r$, and $\delta A_r$ is the area of the sampling region at radial displacement $r$. The normalised number density is then evaluated according to $\hat{\rho} = \rho / \rho_0$, where the reference number density is given by $\rho_0 = N / A$ with $N$ being the total number of particles in the domain and $A$ the total domain area.
In a particle pair context $\hat{\rho}$ can then be interpreted as the RDF, with the concentration build-up around
$r=0$ that arises due to the form of the inhomogeneous velocity field representing a measure of clustering.
Simulations are initialised with a uniform particle distribution and run until the concentration $\hat{\rho}$
equilibrates to a steady state profile before further time integrations are performed to calculate the quantities needed
to compute the statistical ensembles in the integrands of Eqs.~\eqref{eq:dispersion-coefficients}.

Contributions from particle trajectories are accumulated within a \textcolor{black}{radial Eulerian grid}
using a cell-based inverse distance weighted averaging procedure. This accounts for the location of particles within
grid elements, thereby providing $C^1$ continuity and reducing the noise generated from the statistical averaging process compared to a box counting approach ($C^0$ continuity). Additional smoothing of quantities is carried out using a Savitzky-Golay filter with a stencil size of 15 grid points in order to further reduce the statistical noise from simulation data \cite{Savitzky1964}.
The spatial derivatives needed for evaluation of
$\nabla \cdot \overline{\boldsymbol{\lambda}}$ and
$\nabla \cdot \overline{\bm{c}\bm{c}}$
(see Eq.~\eqref{eq:div-lambda}) are also obtained from simulation data using Savitzky-Golay convolution coefficients.

Evaluation of $\tau_{Lp}$ in Eq.~\eqref{eq:lagrangian-integral-timescale} is carried out by calculating $\tau_{L}$ as the integral timescale based upon the fluid velocity correlations seen by particles in the KS flow field.

\section{Results} \label{sec:results}

\subsection{Radial dispersion coefficients and kinetic stresses}

\begin{figure}[!ht]
    \begin{subfigure}[c]{\columnwidth}
        \includegraphics[width=\textwidth,trim={0 0 0 0}]{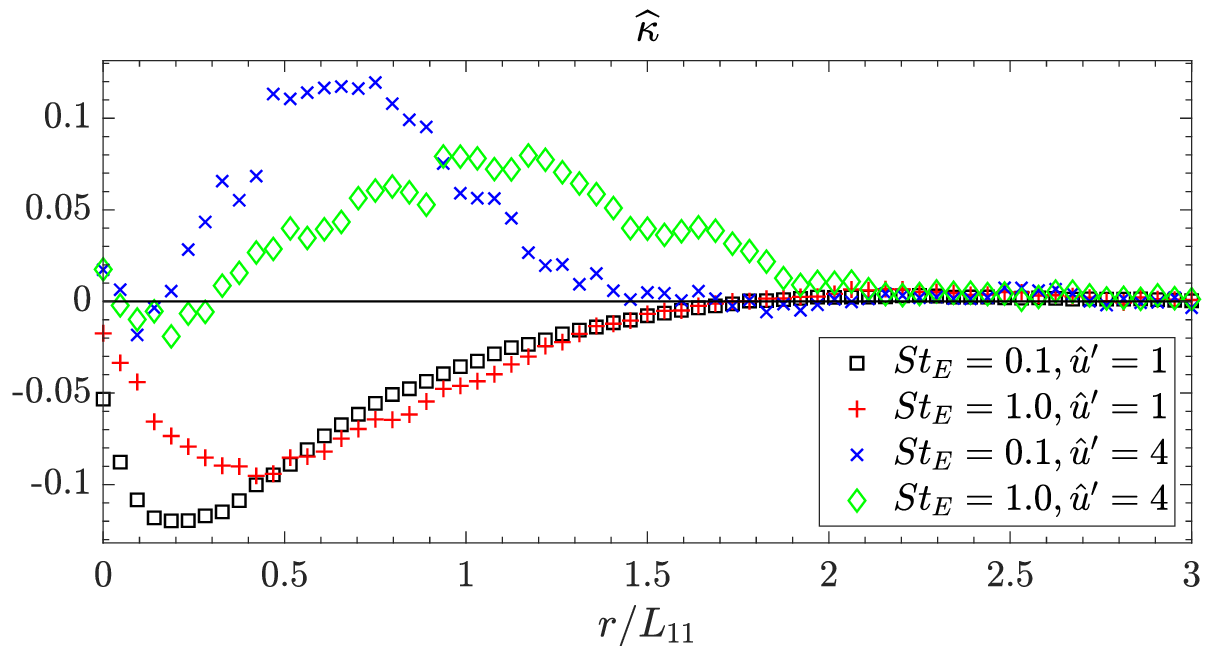}
        \caption{}
        \label{fig:kappa-plot}
    \end{subfigure}
    \begin{subfigure}[c]{\columnwidth}
        \includegraphics[width=\textwidth,trim={0 0 0 0}]{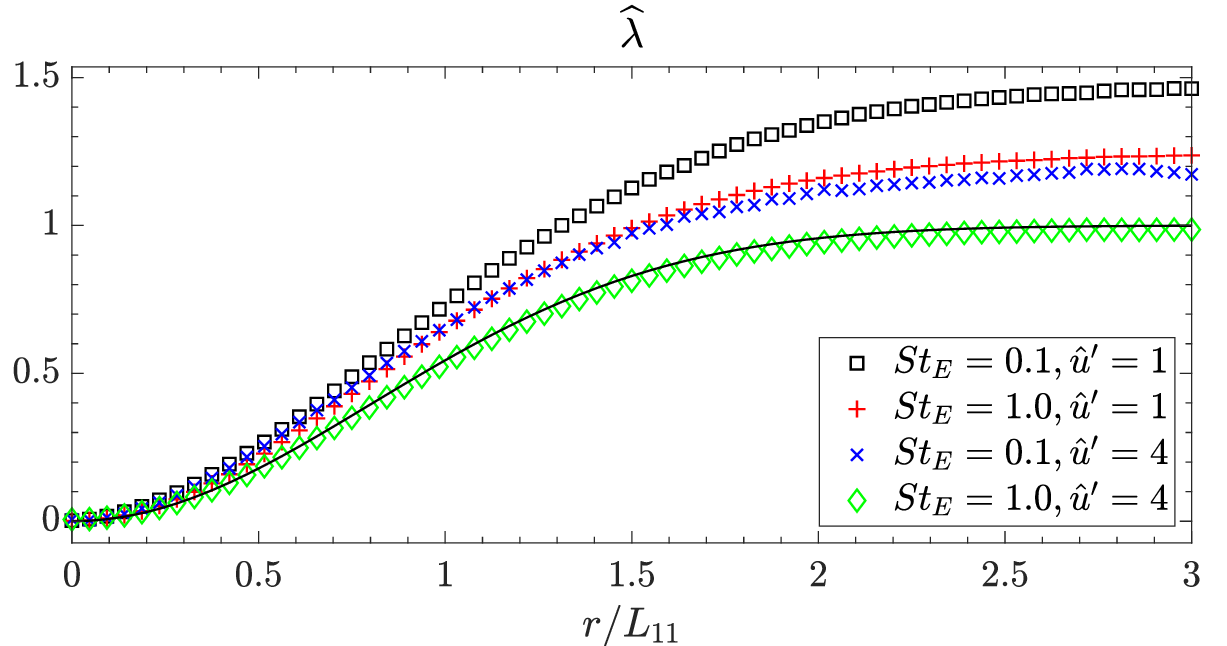}
        \caption{}
        \label{fig:lambda-plot}
    \end{subfigure}
    \begin{subfigure}[c]{\columnwidth}
        \includegraphics[width=\textwidth,trim={0 0 0 0}]{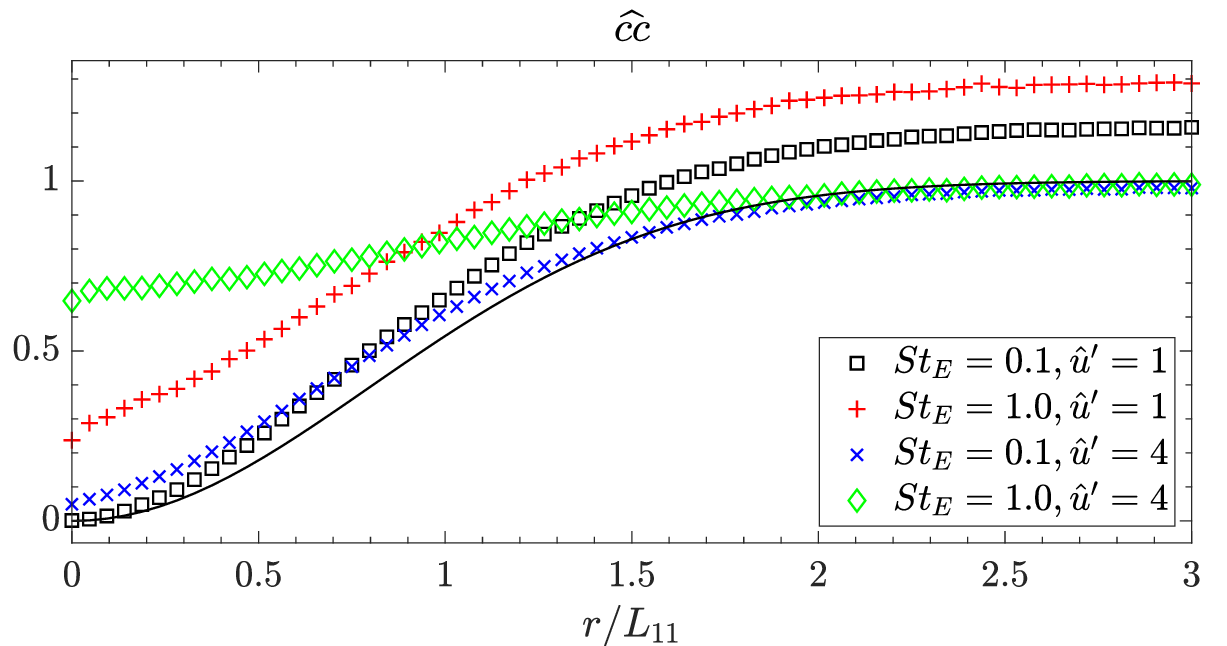}
        \caption{}
        \label{fig:cc-plot}
    \end{subfigure}
    \caption{
        Radial coefficients: \eqref{fig:kappa-plot} $\widehat{\kappa}$;
        \eqref{fig:lambda-plot} $\widehat{\lambda}$;
        \eqref{fig:cc-plot} $\widehat{cc}$. \newline
        The LHA forms $\widehat{\kappa}^{\text{L}}$, $\widehat{\lambda}^{\text{L}}$, $\widehat{cc}^{\text{L}}$ are given by the solid lines.
    }
    \label{fig:disp-cc}
\end{figure}

It is convenient to introduce normalised measures for the various flux terms studied.
This allows for easy comparison of the relative significance of these terms
across a range of different turbulence scales, as characterized by $St_E$ and $\hat{u}^{\prime}$.
To this end scaled coefficients are introduced:
\begin{equation}
\widehat{\kappa} = \overline{\kappa}_{\parallel} / \kappa^{*} 
\, , \quad
\widehat{\lambda} = \overline{\lambda}_{\parallel} / \lambda^{*}
\, , \quad
\widehat{cc} = \overline{cc}_{\parallel} / cc^{*}
\, ,
\end{equation}
where the normalisation factors $\kappa^{*}$, $\lambda^{*}$ and $ cc^{*}$ are
\begin{subequations}
    \label{eq:norm-disp-cc}
    \begin{align}
        \kappa^{*} & = \sigma^2 / ( \tau_{Lp} {u}^\prime )  \, , \label{eq:kappa-norm} \\
        \lambda^{*} = \lim\limits_{r\rightarrow\infty} \overline{\boldsymbol{\lambda}}^{\text{L}} & =
 \sigma^2
        \, , \label{eq:lambda-norm} \\
        cc^{*} = \lim\limits_{r\rightarrow\infty} \overline{\bm{c}\bm{c}}^{\text{L}} & =
  \sigma^2 St_{Lp}
        \, . \label{eq:cc-norm}
    \end{align}
\end{subequations}
The choice of asymptotic LHA expressions for $\lambda^{*}$ and $cc^{*}$ is natural
since this permits straightforward assessment of how the exact, spatially varying values
deviate from these approximations. Further, the corresponding normalized LHA profiles, 
$\widehat{\lambda}^{\text{L}}$ and $\widehat{cc}^{\text{L}}$,  collapse onto single curves,
independent of $St_{E}$ and $\hat{u}^{\prime}$.
The dependence of $\widehat{\kappa}$, $\widehat{\lambda}$ and $\widehat{cc}$ on the normalised
separation distance $r/L_{11}$ 
is illustrated in Fig.~\ref{fig:disp-cc}, which shows steady-state profiles for
a range parameters values, $St_E = 0.1$, $1.0$ and $\hat{u}^{\prime} = 1$, $4$.

A number of important observations emerge from these:
Firstly, Fig.~\eqref{fig:kappa-plot} shows that $\widehat{\kappa}$ is clearly non-zero for all but very large separations.
Again this emphasises the need to include this term in models. Not only those designed to describe particle clustering phenomena but, more generally, particle transport in inhomogeneous turbulence.
Evidently, and consistent with uniformly mixed particles in homogenous turbulence, $\widehat{\kappa} \rightarrow 0$
with increasing separation.
The profiles in Fig.~\eqref{fig:kappa-plot} indicate a strong dependence of $\widehat{\kappa}$ on $St_E$, with the extrema of $\widehat{\kappa}$
being orders of magnitude greater for $St_E = 0.1$ than for $St_E = 1.0$.
Given the form of the scaling factor $\kappa^{*}$ this trend will also be observed in profiles for $\overline{\kappa}_{\parallel}$.
In contrast, for a given value of $St_E$, the different cases of $\hat{u}^{\prime}$ show profiles of similar magnitude.
However, and very significantly, the sense in which the drift embodied in $\widehat{\kappa}$ acts is reversed.
The resulting flux being essentially negative when $\hat{u}^{\prime} = 1$ but positive for the cases where
$\hat{u}^{\prime} = 4$.
Furthermore, the value of separation $r$ at which the peak magnitude of $\widehat{\kappa}$ is observed is greater for larger values of both $St_E$ and $\hat{u}^{\prime}$.
The value of $\widehat{\kappa}$ at $r = 0$ remains nonzero across the different cases. This is an consequence
of the form of the spatial correlation derivative $\nabla \bm{R}$ in Eq.~\eqref{eq:kappa}. Specifically, from the derivative of Eq.~\eqref{eq:inhomogeneous-correlations-decomp} it is clear that $\nabla \bm{R} (\bm{r}^{\prime},t^{\prime}; \bm{0},t)$ is not necessarily zero. It is seen that $\widehat{\kappa} ( 0 )$ is larger for the smaller value of $St_E$.

The behaviour of $\widehat{\lambda}$ is depicted in Fig.~\eqref{fig:lambda-plot}.
The fact that $\widehat{\lambda} (0) = 0$, independent of  both $St_E$ and $\hat{u}^{\prime}$,
follows immediately from Eq.~\eqref{eq:lambda} and the observation that,
from Eq.~\eqref{eq:inhomogeneous-correlations-decomp},
$\bm{R} (\bm{r}^{\prime},t; \bm{0},t) = \bm{0}$.
The variation of $\widehat{\lambda}$ with $r$
mirrors the behaviour of the mean square fluctuating flow velocity profile in Fig.~\ref{fig:RMSI_plot}, with profiles that approach asymptotic values with increasing $r$.
These limiting values being those that would be obtained for a one-particle model in a homogeneous flow. 
The normalized local homogeneous approximation $\widehat{{\lambda}}^{\text{L}}$
is also plotted in Fig.~\eqref{fig:lambda-plot}.
As mentioned, this collapses to a universal profile because of the choice of normalization, and although this approximation is qualitatively representative of the exact profiles it can be seen that there are clear quantitative differences. Even at large separations, in essentially homogeneous regimes, the local homogeneous approximation can fail to accurately predict the value of this dispersion coefficient.

The accuracy of the approximation $\widehat{\lambda}^{\text{L}}$
is seen to vary across the different values of $St_E$ and $\hat{u}^{\prime}$,
with better agreement for the larger values of $St_E$ and $\hat{u}^{\prime}$.
In particular, for the case $St_E = 1.0$, $\hat{u}^{\prime} = 4$, the LHA values are seen to align closely with those
of $\widehat{\lambda}$. There is actually a slight over-estimation.
This can only occur if the model for $\tau_{Lp}$ also gives over-estimated values,
and this will be the case here since the expression given in Eq.~\eqref{eq:lagrangian-integral-timescale}
was proposed using simulation data for which the turbulence structure parameter $\hat{u}^{\prime} = 1$.
Nonetheless, use of the existing model for $\tau_{Lp}$ still captures the qualitative characteristics
of $\widehat{\lambda}$ when $\hat{u}^{\prime} = 4$, and produces a good approximation to the true behaviour.
Clearly the dependence of $\widehat{\lambda}$ on particle separation is influenced
by the form of the longitudinal correlation coefficient $f(r)$ which, in this work, is defined by
Eq.~\eqref{eq:longitudinal-correlation-coefficient}.
The parameter $\sigma_k$ intrinsic to this definition is uniquely determined by $L_{11}$,
and since this is treated as a constant in this study it follows that the profiles of
$\widehat{\lambda}$ is qualitatively similar across all values of $St_E$ and $\hat{u}^{\prime}$.

The kinetic stresses $\widehat{cc}$ are displayed in Fig.~\eqref{fig:cc-plot}.
It is seen that, for low inertia particles $\widehat{cc} ( 0 )  \approx 0$,
whereas, for higher inertia particles this is not the case.
This is what would be expected, and is analogous to the trend observed for particles in turbulent boundary layers.
For smaller particles the kinetic energy associated with particle fluctuations
is more responsive to the near-wall attenuation of turbulent kinetic energy.
Again, the local homogeneous approximation $\widehat{cc}^{\text{L}}$ is seen not to be a uniformly good predictor
of this particle statistic. By construction, from Eqs.~\eqref{eq:LHA-dispersion-coefficients} and \eqref{eq:cc-LHA},
it follows that $\widehat{cc}^{\text{L}}(0) = 0$,
and so the approximation can only provide good estimates at small separations for very low inertia particles.
It is also seen to be a poor predictor in the homogeneous (large separation) regime for
$\hat{u}^{\prime} = 1$ where it significantly under-estimates the true statistics.

In summary, $\widehat{\kappa}$, $\widehat{\lambda}$ and $\widehat{cc}$ all exhibit
marked differences from their corresponding local homogeneous approximations.
The later are, in general, poor models for these key particle statistics.
Not only in the strongly inhomogeneous
regime of small separations, but also in essentially homogeneous flows obtained at large separations.
Given the central role that these coefficients play in the net mass flux balance, and hence in particle clustering,
further analysis is warranted
to assess the relative contribution that each plays. This is considered next.

\subsection{Diffusive and convective mass flux contributions}

This section assesses the contributions of the various flux terms in
Eqs.~\eqref{eq:mass-flux}
to the equilibrium, net zero balance
Eq.~\eqref{eq:zero-flux}.
Again, KS-based particle trajectory simulations are used to evaluate each contribution exactly
from the underlying kinetic theory representations, Eqs.~\eqref{eq:dispersion-coefficients},
avoiding any uncertainties associated with the introduction of closure modelling in these.
Therefore, allowing for the statistical noise intrinsic to sample based ensemble averaging,
these results provide exact measures for these flux components. 

\begin{figure}
    \includegraphics[width=\columnwidth,trim={0 0 0 0}]{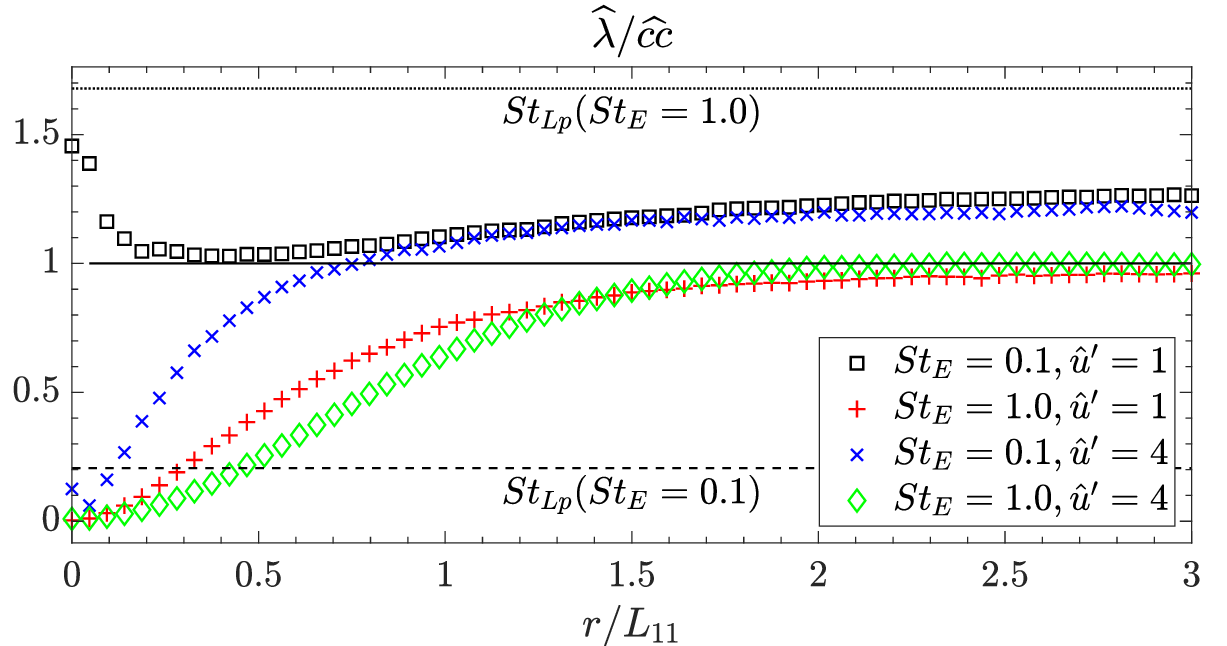}
    \caption{
        Ratio $\widehat{\lambda} / \widehat{cc}$ of normalised diffusive flux components.
        The values of $St_{Lp}$, corresponding to $St_E = 0.1$ and $St_E = 1.0$, are represented by the dashed and dotted lines respectively. 
    }
    \label{fig:diffusive-flux}
\end{figure}

The effective diffusion coefficient $D_{\parallel}$, Eq.~\eqref{eq:D-radial}, depends on the radial components of both
the particle kinetic stresses, $\overline{cc}_{\parallel}$, and the interfacial momentum exchange as 
captured by $\overline{\lambda}_{\parallel}$.
The relative contribution of these two components can be investigated by considering their ratio,
and Fig.~\ref{fig:diffusive-flux} presents plots of the normalized ratio 
\begin{equation}
\frac{\widehat{\lambda}}{\widehat{cc}} =
\frac{\overline{\lambda}_{\parallel}}{\overline{cc}_{\parallel}} \,
St_{Lp} \, .
\end{equation}
The scaling factor $St_{Lp}$ provides a reference value against which the balance of contributions from the unscaled quantities $\overline{\lambda}_\parallel$ and~$\overline{cc}_\parallel$ can be assessed:
if $\widehat{\lambda} / \widehat{cc} > St_{Lp}$ then $\overline{\lambda}_{\parallel}$ is the major contributor.
Conversely, $\widehat{\lambda} / \widehat{cc} < St_{Lp}$ implies that $\overline{cc}_{\parallel}$
is more significant.
Two lines, associated with the reference values $St_{Lp} = 0.21$ (corresponding to $St_E = 0.1$),
and $St_{Lp} = 1.68$ ($St_E = 1.0$) are plotted
in Fig.~\ref{fig:diffusive-flux}, which also shows the constant value of the normalized LHA ratio,
$\widehat{\lambda}^{\text{L}}_{\parallel} / \widehat{cc}^{\text{L}}_{\parallel} = 1$ independent of $St_E$ and~$\hat{u}^{\prime}$.

Several important findings emerge from the results shown in Fig.~\ref{fig:diffusive-flux}.
Firstly, in the homogeneous (large $r$) regime the relative contribution of the two diffusion coefficients,
$\overline{\lambda}_{\parallel}$, $\overline{cc}_{\parallel}$ is seen to be highly dependent upon $St_{E}$.
For $St_{E} = 0.1$ it is $\overline{\lambda}_{\parallel}$ that dominates, this being approximately six times greater in magnitude than
$\overline{cc}_{\parallel}$.
The situation is reversed for $St_{E} = 1.0$, where the value of $\overline{cc}_{\parallel}$ is approximately twice that of 
$\overline{\lambda}_{\parallel}$. Also notable is that these ratios are independent of the homogeneous turbulent field kinetic energy,
with $\overline{\lambda}_{\parallel}$ and $\overline{cc}_{\parallel}$ scaling similarly with $\hat{u}^{\prime}$.
The situation is markedly different in the inhomogeneous region ($r < L_{11}$):
For $St_{E} = 1.0$ the similarity scaling with $\hat{u}^\prime$ still persists, with $\overline{cc}_{\parallel}$ becoming increasingly dominant
over $\overline{\lambda}_{\parallel}$ as $r  \rightarrow 0$. However, for $St_{E} = 0.1$  this scaling breaks down and very different 
balances are observed for $\hat{u}^\prime = 1$ and $\hat{u}^\prime = 4$.

At the smaller kinetic energy value the diffusional contribution of $\overline{\lambda}_{\parallel}$
becomes increasingly dominant as $r \rightarrow 0$, whereas, at the larger energy value, the influence
decreases such that, for $r < 0.1 L_{11}$ the predominant diffusional effect comes from~$\overline{cc}_{\parallel}$.
\begin{figure*}[!ht]
    \begin{subfigure}[c]{0.495\textwidth}
        \includegraphics[width=\textwidth,trim={0 0 0 0}]{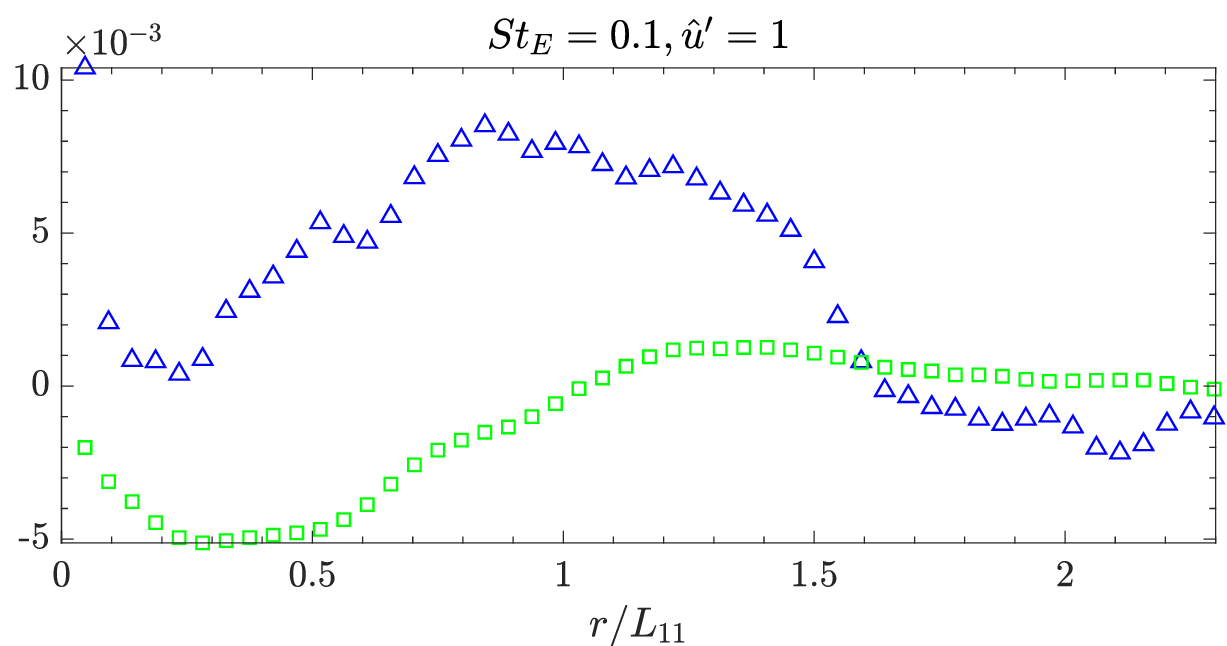}
        \caption{}
        \label{fig:drift-flux_stp1_u1}
    \end{subfigure}
    \begin{subfigure}[c]{0.495\textwidth}
        \includegraphics[width=\textwidth,trim={0 0 0 0}]{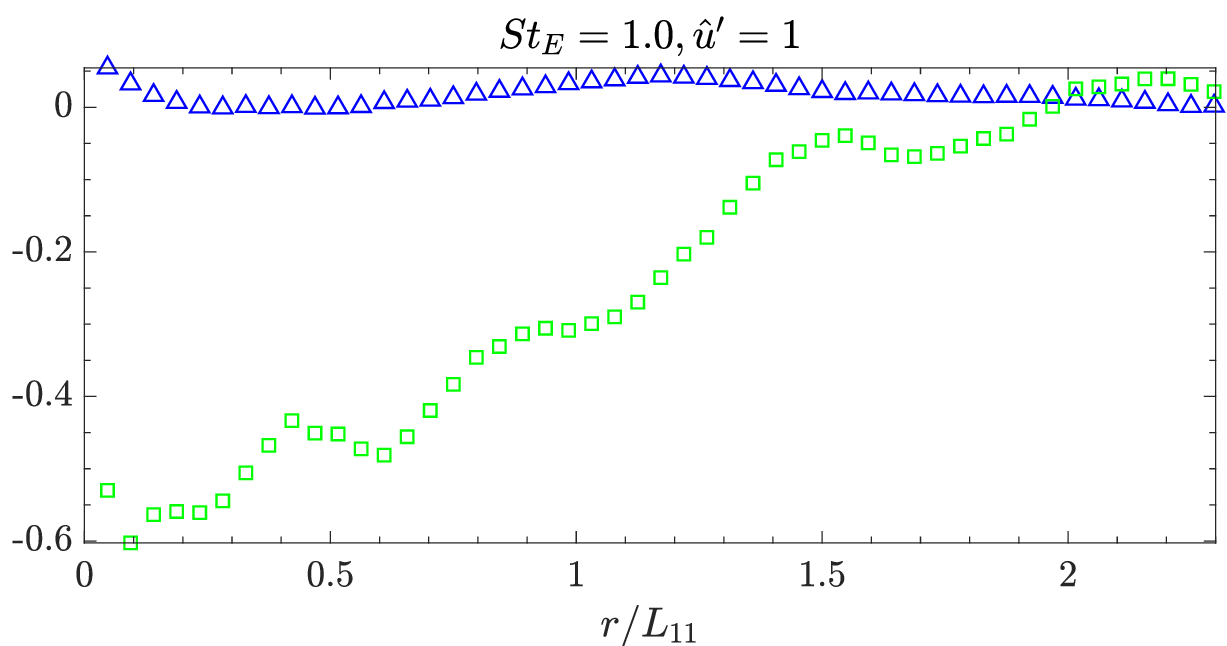}
        \caption{}
        \label{fig:drift-flux_st1_u1}
    \end{subfigure}
    \begin{subfigure}[c]{0.495\textwidth}
        \includegraphics[width=\textwidth,trim={0 0 0 0}]{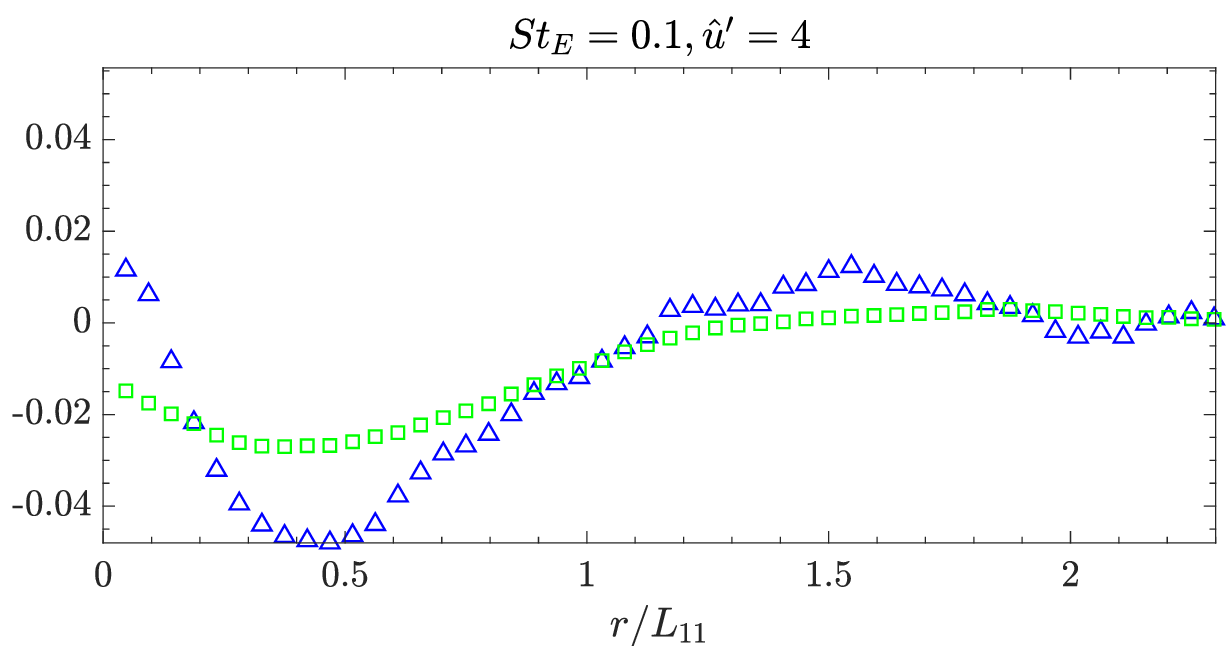}
        \caption{}
        \label{fig:drift-flux_stp1_u4}
    \end{subfigure}
    \begin{subfigure}[c]{0.495\textwidth}
        \includegraphics[width=\textwidth,trim={0 0 0 0}]{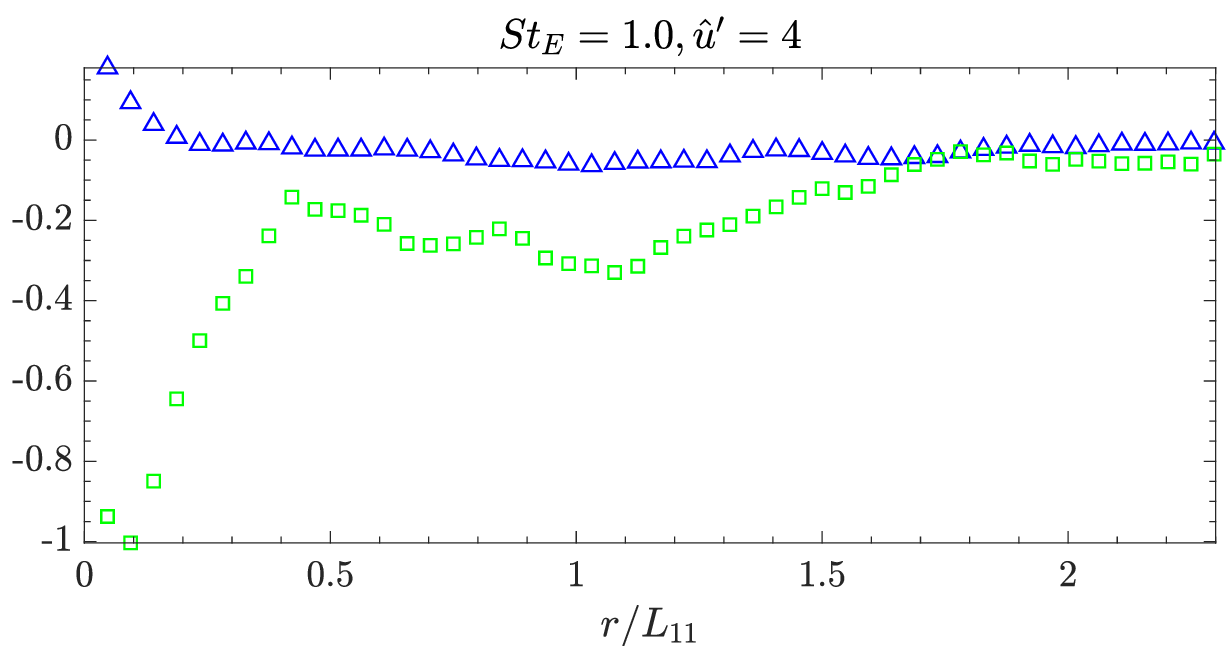}
        \caption{}
        \label{fig:drift-flux_st1_u4}
    \end{subfigure}
    \caption{
        Normalized radial drift flux components:  $\widehat{d}^{\ 1}$ ({\color{blue}$\boldsymbol{\bigtriangleup}$})
and $\widehat{d}^{\ 2}$ ({\color{green}$\boldsymbol{\Box}$})  
    }
    \label{fig:drift-flux}
\end{figure*}

Turning to the convective flux,
the contributions from the drift and turbophoretic terms in
Eq.~\eqref{eq:d-radial} are investigated via the corresponding normalised quantities
\begin{subequations}
\label{eq:d-norm}
\begin{align}
\widehat{d}^{\ 1}
&
=
\left (
\overline{\boldsymbol{\kappa}} - \nabla \cdot
\overline{\boldsymbol{\lambda}}
\right )_{\parallel} / \beta u^{\prime} \, ,
\label{eq:d1-norm}
\\
\widehat{d}^{\ 2}
&
=
-
\left (
\nabla \cdot \overline{\bm{c}\bm{c}}
\right )_{\parallel} / \beta u^{\prime} \, .
\label{eq:d2-norm}
\end{align}
\end{subequations}
The behaviour of these terms is shown in Fig.~\ref{fig:drift-flux}.
As would be expected,
both of these drift velocity contributions become zero in the homogeneous far field.
Therefore, and in contrast to the diffusive terms,
it is more appropriate to present the two contributions $\widehat{d}^{\ 1}$ and
$\widehat{d}^{\ 2}$  separately, rather than in the form of a ratio.
Both terms exhibit distinctly non-zero behaviour at small separations $r$,
with the relative importance of each being determined by both  $St_E$ and $\hat{u}^{\prime}$.

For $St_E = 0.1$,
Figs.~\eqref{fig:drift-flux_stp1_u1} and \eqref{fig:drift-flux_stp1_u4}
show that the non-local turbulent drift $\widehat{d}^{\ 1}$ plays a dominant role in particle clustering.
Strikingly, this flux is seen to act in different directions, depending
on the turbulence intensity:
being positive for $\hat{u}^{\prime} = 1$, Fig.~\eqref{fig:drift-flux_stp1_u1},
but negative for \mbox{$\hat{u}^{\prime} = 4$}, Fig.~\eqref{fig:drift-flux_stp1_u4}.
Consequently, for $\hat{u}^{\prime} = 1$ the net convective flux $\widehat{d}^{\ 1} +\widehat{d}^{\ 2}$ is actually positive in the region \mbox{$0.5 < r/L_{11}  < 1.5$}, reducing the propensity of particles to  cluster.
Only for $r/L_{11} < 0.5$ is the net convective flux negative, with a maximal magnitude attained at $r/L_{11} \approx 0.2$.
For $St_E = 1.0$ the situation is markedly different, as depicted in Figs.~\eqref{fig:drift-flux_st1_u1} and~\eqref{fig:drift-flux_st1_u4}.
The turbophoretic drift $\widehat{d}^{\ 2}$
is now the principle driver of clustering, being an order of magnitude greater than $\widehat{d}^{\ 1}$ at $r = 0$.

The variation of the non-local turbulent drift
$\widehat{d}^{\ 1}$
with $\hat{u}^{\prime}$ can be attributed to the balance between the radial and centrifugal terms in the the divergence
$(\nabla \cdot \overline{\boldsymbol{\lambda}} )_{\parallel}$ given by Eq.~\eqref{eq:div-lambda}. This determines the overall sign of
$( \nabla \cdot \overline{\boldsymbol{\lambda}} )_{\parallel}$,
which is found to be negative for $\hat{u}^{\prime} = 1$ and positive for $\hat{u}^{\prime} = 4$.
This effect follows the same trend as that exhibited by
$\overline{\kappa}_{\parallel}$ in Fig.~\eqref{fig:kappa-plot},
however it has been observed that
$\lvert ( \nabla \cdot \overline{\boldsymbol{\lambda}} )_{\parallel} \rvert > \lvert \overline{\kappa}_{\parallel} \rvert$
in all cases. Thus when
$\overline{\kappa}_{\parallel} > 0$, it follows that 
$\widehat{d}^{\ 1} < 0$, and \textit{vice-versa}.

Eq.~\eqref{eq:RDF} shows how the convective flux, $d_{\parallel}$, and the diffusion coefficient, $D_{\parallel}$,
govern the particle concentration profile,
with a larger convective flux 
and a smaller diffusion coefficient
acting to increase the particle concentration.
Therefore, from Fig.~\ref{fig:drift-flux}, it is anticipate that the preferential particle concentration
will be greatest for $St_E = 0.1$ and $\hat{u}^{\prime} = 4$,
which is indeed the case (see later, Fig.~\ref{fig:numden-plot}).
Since the non-local turbulent drift
$\widehat{d}^{\ 1}$
is dominant in this case, it is therefore this quantity that is principally responsible for the
marked increase in particle concentration at $r = 0$.
Similarly,  the concentration increase for $St_E = 1.0$ and $\hat{u}^{\prime} = 1$ is mainly attributable
to the turbophoretic contribution $\widehat{d}^{\ 2}$ that is then dominant.
Thus there are two separate mechanisms associated with
preferential particle concentration.  Further, in this regard, these terms can be complimentary or
antagonistic, depending upon the values of $St_E$ and $\hat{u}^{\prime}$.

\subsection{Radial distribution function}

\begin{figure}[!ht]
    \includegraphics[width=\columnwidth,trim={0 0 0 0}]{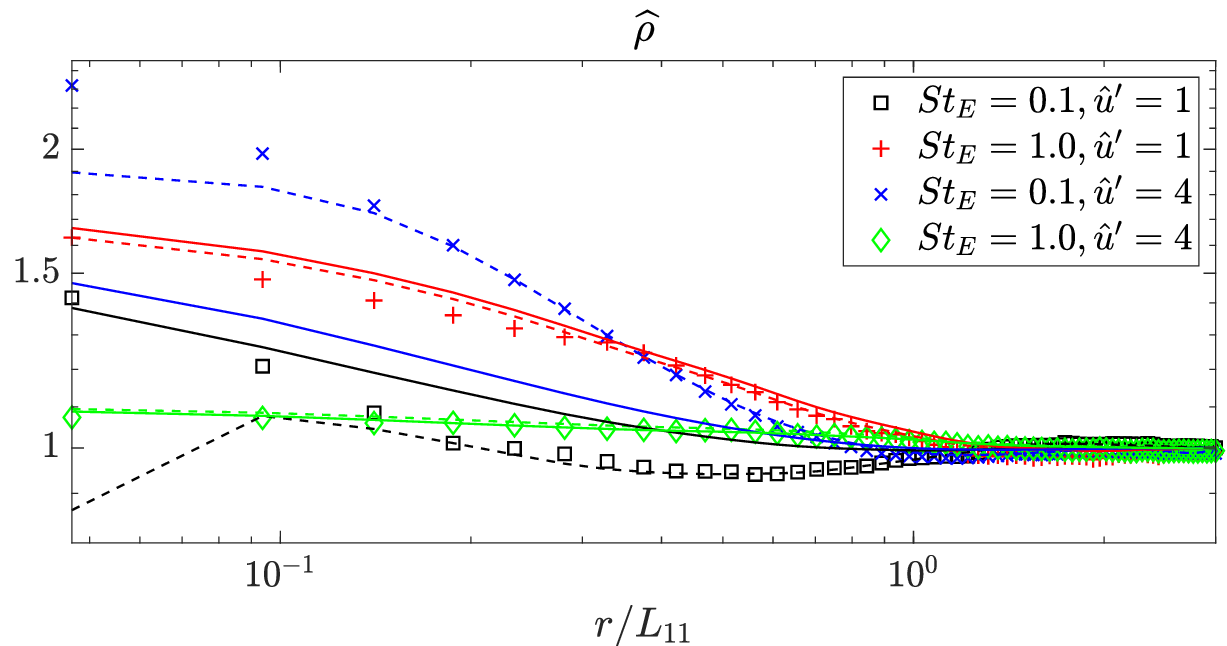}
    \caption{
        Radial distribution function $\widehat{\rho}$. \newline
        Benchmark data (particle counting) shown by markers. \newline
        Solutions given by Eq.~\eqref{eq:RDF} shown by dashed lines. \newline
       Solutions omitting drift $\widehat{d}^{\ 1}$ shown by solid lines.
    }
    \label{fig:numden-plot}
\end{figure}

The normalised RDF $\widehat{\rho} = \rho / \rho_\infty$ provides a natural measure of particle clustering,
and the kinetic theory based representation of this function allows for detailed scrutiny of how
the various flux contributions affect clustering.
Profiles of the RDF, for the different values of $St_{E}$ and $\hat{u}^{\prime}$, are shown in Fig.~\ref{fig:numden-plot}.
These profiles are computed from Eq.~\eqref{eq:RDF}, making use of Eqs.~\eqref{eq:d-D-radial}
and the underlying formulae given by Eqs.~\eqref{eq:dispersion-coefficients}.
To validate these results (and, \textit{a fortiori}, the formulae) Fig.~\ref{fig:numden-plot} also shows values of $\widehat{\rho}$
computed directly, via Eq.~\eqref{eq:number-density-calculation}, using particle trajectory simulations.
As previously mentioned, the drift contribution ${d}^{1}_{\parallel}$ is often overlooked in the
formulation and application of transport equations for disperse particle flows.
It is therefore important to ascertain to what extent such
omission might compromise the predictive capabilities of resulting models.
To this end Fig.~\ref{fig:numden-plot}
also presents the corresponding RDF profiles obtained when this drift term is omitted.

Comparison of the kinetic theory derived profiles with the data obtained from simple particle counting demonstrates the veracity of the PDF representation.
Some discrepancies are observed for $r \ll L_{11}$, especially in the low inertia cases, $St_{E} = 0.1$.
These differences are attributable to numerical error associated from the large flux gradients
present at small separations; the relatively low particle seeding ($10^6$) does not admit high fidelity resolutions of
these gradients based on the underlying ensemble  averaging.
Notwithstanding this, Fig.~\ref{fig:numden-plot} still serves to highlight the ability of the
mass flux terms provided by the kinetic approach to represent the concentration profiles
across the entire parameter space, and offers a verification of Eq.~\eqref{eq:RDF}.

The RDF profile is seen  to be sensitive to changes in both $St_E$ and $\hat{u}^{\prime}$.
As expected, for the lower inertia particles (\mbox{$St_E = 0.1$}) there is more pronounced clustering,
with larger gradients than for $St_E = 1.0$.
It is also notable that, for the case of $St_E = 0.1$ and $\hat{u}^{\prime} = 1$
there is a mid-range separation region in which
$\widehat{\rho}$ actually attains values below the far-field normalized unit value.

At zero separation the highest concentration is obtained with $St_E = 0.1$ and $\hat{u}^{\prime} = 4$, whilst the lowest concentration occurs with $St_{E} = 1.0$ and $\hat{u}^{\prime} = 4$.
The higher level of turbulence intensity therefore has a notable effect on the peak concentration of particles across different values of $St_E$.
By comparison, for $\hat{u}^{\prime} = 1$ the concentration increase at $r=0$ is of a similar size for both $St_E = 0.1$ and $St_E = 1.0$,
demonstrating that the lower value of $\hat{u}^{\prime}$ has less of an effect on the peak concentration level.
This is consistent with previous work \cite{Bragg2014a} which determined that peak clustering occurs at $St_\eta \approx 1$, below which preferential concentration diminishes regardless of the value of $\hat{u}^{\prime}$ as the particle inertia tends to that of fluid tracers.

The solid lines in Fig.~\ref{fig:numden-plot}
show the profiles obtained from Eq.~\eqref{eq:RDF} but with the drift flux contribution, ${d}^{1}_{\parallel}$, omitted.
For $St_E = 1.0$ these profiles are seen to align closely with the exact forms (dashed lines).
This reflects the fact that for high inertia particles the turbophoretic flux,
${d}^{2}_{\parallel}$, is the dominant contribution to particle clustering,
the non-local turbulent drift ${d}^{1}_{\parallel}$ being negligible in comparison.

In stark contrast, for $St_E = 0.1$, the concentration profiles obtained are markedly different.
This emphasises the central importance of including this drift flux contribution, not only in models
for particle clustering but, more generally, in dispersion models for particles in inhomogeneous flow regimes.

\section{Conclusions} \label{sec:conclusions}

The distribution and dynamics of monodisperse particles in inhomogeneous turbulence has been considered, making use of a PDF representation formally derived in the framework of a kinetic theory. This allows for the precise and detailed formulation of various key terms that contribute to the local particle mass flux.

These terms have been computed in the context of a particle pair dynamics, where they govern the degree of particle clustering that emerges. The construction of an inhomogeneous Gaussian flow field mimicking that associated with the relative dynamics of particle-pairs ensures that the resulting flux representations are exact, so that the significance of the various flux terms can be assessed without the need of potentially confounding closure approximations. The inhomogeneity of the flow can be considered analogous to that in one-particle turbulent boundary layer modelling used to study particle deposition {\color{corr}within channel and pipe flow\cite{VanDijk2012}, with the balance of convective and diffusive contributions to the particle mass flux then determining the particle concentration in a given region}.

The results show that the dominant contribution in both the diffusive and the convective fluxes is determined principally by the Stokes number $St_E$, with the turbulence intensity $\hat{u}^{\prime}$ generally playing a lesser role.
In the diffusive flux the PDF dispersion tensor $\overline{\lambda}_{\parallel}$ is dominant for small $St_E$, whereas the kinetic stresses $\overline{cc}_{\parallel}$ dominate for large $St_E$.

In the convective flux  the non-local turbulent drift velocity $d^{1}_{\parallel}$ is the dominant contribution at small $St_E$, whilst the turbophoretic flux component $d^{2}_{\parallel}$ dominates for large $St_E$.
The turbophoretic flux $d^{2}_{\parallel}$ is negative across all values of $St_E$ and $\hat{u}^{\prime}$, and therefore always acts to promote particle clustering.
However the sign of $d^{1}_{\parallel}$ varies depending upon $\hat{u}^{\prime}$ and, due to the contributions determined by ${\kappa}_{\parallel}$ and $(\nabla \cdot \bm{\lambda} )_{\parallel}$, is only negative for small $St_E$ and large $\hat{u}^{\prime}$.
Since the net convective velocity is given by the sum ${d}^{1}_{\parallel} + {d}^{2}_{\parallel}$ it follows that particle clustering is most pronounced for small $St_E$ and large $\hat{u}^{\prime}$.
In contrast ${d}^{1}_{\parallel}$ acts to decrease the preferential concentration of particles when positive.

The results serve to demonstrate that the PDF-based representation of the flux contributions faithfully captures the influence of  inhomogeneities in the turbulence.
In particular the work establishes that, in the context of particle clustering, there are two distinct and sometimes competing mechanisms that influence the build-up of particle concentration. This holds implications for particle transport in bidisperse systems, with a view to improved predictions of agglomeration and the particle collision kernel \cite{Reeks2021}, and would enable particle growth to be considered as an evolution of the particle size distribution.

Closure modelling in previous works has frequently invoked approximations that, in essence, equate to
$\overline{\boldsymbol{\kappa}} - \nabla \cdot \overline{\boldsymbol{\lambda}} \approx \bm{0}$.
The work presented here demonstrates that, in general, there are no grounds for this simplification, and that care is needed in dealing with these terms. Notably, within the context of clustering, it is shown that neglecting this contribution results in significant errors in the build-up of particle concentration. This highlights the importance of the kinetic approach in its ability to deal with effects due to both the large and small scales of the turbulence, and in particular at identifying the convective and diffusive contributions within the particle mass flux balance which are responsible for this behaviour. This provides the motive for developing improved closures for the non-local turbulent drift flux $\overline{\boldsymbol{\kappa}} - \nabla \cdot \overline{\boldsymbol{\lambda}}$ in future work. This is of relevance for two-fluid modelling of dispersed particle flows in turbulence, which whilst being representative of a range of industrial and environment flow scenarios, is limited in its applicability by existing closures that do not capture the effect of turbulence on the particle phase well.

\begin{acknowledgments}
The authors acknowledge the Ph.D. funding from the School of Engineering, Newcastle University, to support C.P.S.’s research, and also the Rocket High Performance Computing service at Newcastle University.
\end{acknowledgments}

\section*{Author Declarations}

The authors have no conflicts to disclose.

\section*{Data Availability Statement}
The data that support the findings of this study are available
from the corresponding author upon reasonable request.

\appendix

{\color{corr}
\section{Derivation of the particle mass flux balance} \label{sec:mass-flux-balance-derivation}

The particle mass flux balance in Eq.~\eqref{eq:mass-flux} is central to the analysis in this paper, and is obtained from the particle phase momentum equation \eqref{eq:momentum-equation}, repeated here for clarity
\begin{equation} \label{eq:ppme}
	 \rho \frac{D \overline{\bm{w}}}{D t} = - \nabla \cdot \left[ \rho (\overline{\bm{c}\bm{c}} + \overline{\boldsymbol{\lambda}}) \right] + \rho (\overline{\bm{F}} + \overline{\boldsymbol{\kappa}}) \, .
\end{equation}
The density weighted mean force per unit mass acting on particles is given by averaging Eq.~\eqref{eq:mean-force} with the arguments replaced by the two-particle variables
\begin{equation} \label{eq:mean-force-two-particle}
	\overline{\bm{F}} = \beta (\langle \bm{u} \rangle - \overline{\bm{w}})
\end{equation}
Expanding the gradient term in Eq.~\eqref{eq:ppme} produces both convective and diffusive contributions to the terms involving the kinetic stresses $\overline{\bm{c}\bm{c}}$ and kinetic dispersion tensor $\overline{\boldsymbol{\lambda}}$
\begin{equation} \label{eq:gradient-expansion}
	\nabla \cdot \left[ \rho ( \overline{\bm{c}\bm{c}} + \overline{\boldsymbol{\lambda}} ) \right]  = \rho \nabla \cdot \left( \overline{\bm{c}\bm{c}} + \overline{\boldsymbol{\lambda}} \right) + \left( \overline{\bm{c}\bm{c}} + \overline{\boldsymbol{\lambda}}^{\top} \right) \cdot \nabla \rho
\end{equation}
Note that the convective contributions are weighted by $\rho$, whereas the diffusive contributions are weighted by $\nabla \rho$. Substitution of Eqs.~\eqref{eq:mean-force-two-particle} and \eqref{eq:gradient-expansion} into Eq.~\eqref{eq:ppme}, followed by some rearrangement of terms and noting that $\beta = 1 / \tau_p$, results in the particle mass-flux balance
\begin{align}
	\rho \overline{\bm{w}}
	& = \rho \Bigg[
		\left\langle \bm{u} \right\rangle
		+ \tau_p \Bigg\{
		\left[ \overline{\boldsymbol{\kappa}} - \nabla \cdot \overline{\boldsymbol{\lambda}} \right]
		- \nabla \cdot \overline{\bm{c}\bm{c}}
		- \frac{D \overline{\bm{w}}}{D t}  \Bigg\} \Bigg] \nonumber \\
	& - \tau_p \left( \overline{\bm{c}\bm{c}} + \overline{\boldsymbol{\lambda}}^{\top} \right) \cdot \nabla \rho
	\, ,
\end{align}
which is the convective-diffusive representation of the particle phase momentum equation for a linear drag model, and is the form given in Eq.~\eqref{eq:mass-flux}.
}

\section{Kinematic simulation} \label{sec:KS}

The homogeneous isotropic velocity field $\bm{U}(\bm{r},t)$ used as a basis for
constructing associated inhomogeneous flow fields is generated using the technique of kinematic simulation.
The formulation is designed to produce a synthetic turbulent flow with the desired properties.
Specifically, the velocity field is expressed in terms of its Fourier series representation
\begin{equation} \label{eq:KS-flow-field}
    \bm{U}(\bm{r},t) = \sum_{\bm{k}} \bm{c}_{\bm{k}} (t) \exp [i \bm{k} \cdot \bm{r}] \, ,
\end{equation}
where the modes, $\bm{k}$, and coefficients, $\bm{c}_{\bm{k}}$, are prescribed so that the properties of incompressibility, homogeneity, isotropy, and statistical stationarity are all satisfied.
Moreover, the flow field is constructed {\color{corr} using a Gaussian form of longitudinal correlation coefficient
\begin{equation} \label{eq:longitudinal-correlation-coefficient}
	f(r) = \exp \left[ -\frac{1}{2} \sigma_k^2 r^2 \right] \, ,
\end{equation}
making it is consistent with the Batchelor-Townsend energy spectrum \cite{Batchelor1948a}. For the case of $n = 2$ used in this work, the corresponding kinetic energy spectrum $E(k)$ is given by \cite{Kraichnan1970,Schumann1978}
\begin{equation} \label{eq:energy-spectrum}
	E(k) = 2 \pi {u^{\prime}}^2 \frac{k^3}{\sigma_k^4} \exp \left[ -\frac{1}{2} \frac{k^2}{\sigma_k^2} \right] \, ,
\end{equation}
where $k = \lvert \bm{k} \rvert$, and $\sigma_k$ is the magnitude of the modes corresponding to the peak of the energy spectrum $E(k)$, and therefore acts as a measure of the velocity field spatial correlations which are characteristic of the larger energetic eddies.}
Further details of the method are provided in previous work \cite{Stafford2021}.

The form of the velocity field generated by Eq.~\eqref{eq:KS-flow-field}
also permits inferences about the
statistical properties of the {\color{corr} inhomogeneous} effective relative velocity field $\bm{u}(\bm{r},t)$ defined by Eq.~\eqref{eq:inhomogeneous-flu-vel}.
Notably, the specification of $\bm{U}(\bm{r},t)$ using Eq.~\eqref{eq:KS-flow-field} implies that 
\begin{equation} \label{eq:ks-inhomogeneous}
    \bm{u} (\bm{r},t) = \sum_{\bm{k}} \bm{c}_{\bm{k}} (t) \big[ \exp [ i \bm{k} \cdot \bm{r} ] - 1 \big] \, .
\end{equation}
Since the only modification required is a constant shift of the exponential terms in the sum,
it follows that $\bm{u}$ is no more computationally demanding to construct than $\bm{U}$.
Further, generating $\bm{U}$ such that it conforms to a Gaussian field ensures that $\bm{u}$ is also Gaussian.
This important fact is particularly relevant in the context of the kinetic theory used in this work to construct
PDF transport equations for particle dynamics. These equations have been shown to provide exact
descriptions of the distribution dynamics, with the only proviso being that the underlying fluid velocity field is Gaussian.

\bibliography{pp-01.bib}

\end{document}